\documentclass[twocolumn]{aastex701}

\begin{document}

\title{The Evolution of Low-Ionization Electron Densities in Galaxies Across Cosmic Epochs}

\correspondingauthor{M.\ G.\ Stephenson}

\author[0000-0003-4717-0376]{Mabel G. Stephenson}
\affiliation{Department of Astronomy, The University of Texas at Austin, 2515 Speedway Boulevard, Austin, TX 78712, USA}
\affiliation{Cosmic Frontier Center, The University of Texas at Austin, Austin, TX 78712, USA}
\email[show]{msteph@utexas.edu}

\author[0000-0002-0302-2577]{John Chisholm}
\affiliation{Department of Astronomy, The University of Texas at Austin, 2515 Speedway Boulevard, Austin, TX 78712, USA}
\affiliation{Cosmic Frontier Center, The University of Texas at Austin, Austin, TX 78712, USA}
\email{chisholm@austin.utexas.edu}

\author[0000-0001-8419-3062]{A. Saldana-Lopez}
\affiliation{Department of Astronomy, Oskar Klein Centre, Stockholm University, 106 91 Stockholm, Sweden}
\email{alberto.saldana-lopez@astro.su.se}

\author[0000-0003-2871-127X]{Jorryt Matthee}
\affiliation{Institute of Science and Technology Austria (ISTA), Am Campus 1, 3400 Klosterneuburg, Austria}
\email{jorryt.matthee@ist.ac.at}

\author[0000-0002-6586-4446]{Alaina Henry}
\affiliation{Space Telescope Science Institute, 3700 San Martin Drive, Baltimore, MD 21218, USA}
\email{ahenry@stsci.edu}

\author[0000-0003-3997-5705]{Rohan P. Naidu}
\affiliation{MIT Kavli Institute for Astrophysics and Space Research, 77 Massachusetts Avenue, Cambridge, MA 02139, USA}
\email{rnaidu@mit.edu}

\author[0000-0002-0159-2613]{Sophia R. Flury}
\affiliation{Institute for Astronomy, University of Edinburgh, Royal Observatory, Edinburgh, EH9 3HJ, UK}
\email{sflury@roe.ac.uk}

\author[0000-0001-6251-4988]{Taylor A.\ Hutchison}
\altaffiliation{NASA Postdoctoral Fellow}
\affiliation{Astrophysics Science Division, Code 660, NASA Goddard Space Flight Center, 8800 Greenbelt Rd., Greenbelt, MD 20771, USA}
\affiliation{Department of Astronomy, University of Maryland, Baltimore Country, MD 21250, USA}
\affiliation{Center for Research and Exploration in Space Science and Technology, NASA/GSFC, Greenbelt, MD 20771 USA}
\email{astro.hutchison@gmail.com}

\author[0000-0001-8519-1130]{Steven L. Finkelstein}
\affiliation{Department of Astronomy, The University of Texas at Austin, 2515 Speedway Boulevard, Austin, TX 78712, USA}
\affiliation{Cosmic Frontier Center, The University of Texas at Austin, Austin, TX 78712, USA}
\email{stevenf@astro.as.utexas.edu}

\author[0000-0002-7938-614X]{Natalia G. Guseva}
\affiliation{Bogolyubov Institute for Theoretical Physics, National Academy of Sciences of Ukraine, 14-b
Metrolohichna str., Kyiv 03143, Ukraine}
\email{nguseva@bitp.kyiv.ua}

\author[0000-0002-1416-6082]{Yuri I. Izotov}
\affiliation{Bogolyubov Institute for Theoretical Physics, National Academy of Sciences of Ukraine, 14-b
Metrolohichna str., Kyiv 03143, Ukraine}
\email{yizotov@bitp.kyiv.ua}

\author[0000-0001-8442-1846]{Rui Marques-Chaves}
\affiliation{Department of Astronomy, University of Geneva, Chemin Pegasi 51, 1290 Versoix, Switzerland}
\email{Rui.MarquesCoelhoChaves@unige.ch}

\author[0000-0001-5851-6649]{Pascal A. Oesch}
\affiliation{Department of Astronomy, University of Geneva, Chemin Pegasi 51, 1290 Versoix, Switzerland}
\affiliation{Cosmic DAWN Center, Niels Bohr Institute, University of Copenhagen, Jagtvej 128, K\o benhavn N, DK-2200, Denmark}
\email{pascal.oesch@unige.ch}

\author[0000-0001-9687-4973]{Naveen A. Reddy}
\affiliation{Department of Physics and Astronomy, University of California,
Riverside, 900 University Avenue, Riverside, CA 92521, USA}
\email{naveenr@ucr.edu}

\author[0000-0001-7144-7182]{Daniel Schaerer}
\affiliation{Department of Astronomy, University of Geneva, Chemin Pegasi 51, 1290 Versoix, Switzerland}
\affiliation{CNRS, IRAP, 14 Avenue E. Belin, 31400 Toulouse, France}
\email{daniel.schaerer@unige.ch}

\begin{abstract}

We investigate the low-ionization electron density in the interstellar medium of galaxies across $0$$<$$z$$<$$8$, derived with the [\ion{O}{2}]$\lambda\lambda3727,3730$ and [\ion{S}{2}]$\lambda\lambda6718,6733$ doublets, using both deep \textit{JWST} NIRCam/grism and NIRSpec/MSA spectroscopy. Along with ancillary density samples from the literature at $0$$<$$z$$<$$6$, we assemble three samples containing a total of 703 galaxies spanning $3.7$$<$$z$$<$$7.8$ at $\langle z\rangle\sim5.6$; 691 galaxies in the luminosity-complete NIRCam/grism programs CONGRESS and FRESCO at $\langle z\rangle \sim4.3$ and $\langle z\rangle\sim5.3$, respectively, and 12 galaxies with high-resolution NIRSpec/MSA spectra from the GO 1871 and GLASS programs. We find median electron densities of $n_{\rm{e}}$[\ion{S}{2}]$=$$463^{+281}_{-203}~{\rm{cm}}^{-3}$ for CONGRESS, $n_{\rm{e}}$[\ion{S}{2}]$=$$301^{+388}_{-206}~{\rm{cm}}^{-3}$ for FRESCO, and $n_{\rm{e}}$[\ion{O}{2}]$=$$202^{+145}_{-102}~{\rm{cm}}^{-3}$ for the NIRSpec/MSA sample. The median low-ionization $n_{\rm{e}}$ increases from $z= 0$ to $z=4$, in agreement with previous studies and those using \textit{JWST} observations. However, beyond $z\sim4$ median $n_{\rm{e}}$ begins to deviate from a monotonically increasing evolution with $z$. We find a positive, but shallow, correlation between O$_{32}$ (a proxy for the ionization parameter) and $n_{\rm{e}}$, while also confirming a positive correlation between O$_{32}$ and $z$. We suggest that these trends are possibly due to an evolving ISM gas density and ionization structure, and propose a toy model where the gas clumping evolves with $z$ to explain the observed density and ionization variations.

\end{abstract}

% \keywords{}

\section{Introduction} 
\label{sec:intro}

Galaxies evolve hierarchically, beginning with merging primordial dark matter halos that continue to accrete baryons \citep{white_rees_1978}. The first generations of stars form, which in turn produce ionizing photons and synthesize metals. These components then grow and merge to form the earliest galaxies \citep[e.g.][]{bromm_larson_2004, bromm_yoshida_2011}. Extensive observations of galaxies in the local universe ($z\sim0$) provide an expansive view of present-day galaxy astrophysics, but galaxy formation at high-redshift ($z>4$) long remained beyond the reach and scope of observations.

Emission lines reflect the chemical enrichment, dust attenuation, and ionization structure of the interstellar medium (ISM) within their host galaxies \citep{kewley_etal_2019}. One of the most crucial parameters to set emission line strengths is the electron density ($n_{\rm{e}}$), because collisionally-excited emission line strengths depend on $n^2_{\rm{e}}$ (if $n_{\rm{e}}$ is less than the critical density\footnote{The critical density is the density at which the rate of radiative de-excitation equals the rate of collisional de-excitation \citep[e.g.][]{osterbrock_1989}.}, $n_{\rm{crit}}$,  of a given transition), assuming that the $n_{\rm{ion}}/n_{\rm{H}}$ abundance remains constant such that $n^2_{\rm{e}} \approx n_{\rm{e}}n_{\rm{ion}}$. Emission lines, and thus $n_{\rm e}$, also set the phase structure of the ISM, by balancing heating and cooling. Therefore, $n_{\rm{e}}$ is influenced by physical and dynamical processes driven by star formation, e.g. stellar feedback and winds, inflows/outflows of gas, turbulence, shocks, and supernovae \citep{osterbrock_1989, osterbrock_ferland_2006, draine_2011, kewley_etal_2019}.

The electron density is also crucial to understanding the degree of ionization of gas surrounding an ionizing source, which is set by the competition of hydrogen-ionizing ($E_{\rm{IP}}\geq13.60$ eV) photons and recombinations \citep{osterbrock_ferland_2006, draine_2011, sanders_etal_2016}. This degree can be quantified by the dimensionless ionization parameter ($U$, hereafter referred to as the ionization parameter), which is set by the ratio between the ionizing photon density ($n_\gamma$) and the hydrogen density ($n_{\rm H}$);
\begin{equation}
    U\equiv\frac{n_\gamma}{n_{\rm{H}}}.
\label{eq:1}
\end{equation}
In a radiation-bounded \ion{H}{2} region (i.e. no ionizing photons escape), it is often useful to define $U$ in terms of the canonical Strömgren spherical radius ($R_S$), which is the distance where recombination and ionization of H is in equilibrium, assuming a constant density and uniformly filled volume. Galaxies are ensembles of many \ion{H}{2} regions, but are oftentimes approximated as one single Strömgren sphere \citep[e.g.][]{kewley_etal_2019, davies_etal_2021}. If we also assume that the density of hydrogen gas inside the region is equivalent to the electron density in a fully-ionized plasma,
\begin{equation}
    U=\frac{Q}{4\pi R^2_Scn_{\rm{e}}}.
\label{eq:2}
\end{equation}
Using the definition of the Strömgren radius \citep[see e.g.][]{sanders_etal_2016}, we can then simplify the ionization parameter to
\begin{equation}
    U\propto n_{\rm{e}}^{1/3} Q^{1/3} \epsilon^{2/3},
\label{eq:3}
\end{equation}
where the primary dependencies are the electron density ($n_{\rm{e}}$), the rate of ionizing photons ($Q$), and the volume filling fraction ($\epsilon$). However, we stress that these simplifying assumptions do not account for crucial geometric effects present in real \ion{H}{2} regions and galaxies, which have departures from an idealized Strömgren geometry \citep[see][for more information and a more complete derivation of $U$]{charlot_longhetti_2001, sanders_etal_2016}.

The ionization parameter can be approximated observationally in galaxies using strong-line ratios between ionized species of the same element, such as the O$_{32}$ ratio \citep[e.g.][]{kewley_etal_2019, reddy_etal_2023b, reddy_etal_2023a}. Galaxies at higher redshifts are found to have higher ionization parameters at fixed stellar mass \citep[e.g.][]{brinchmann_etal_2008, steidel_etal_2014, sanders_etal_2016, papovich_etal_2022, reddy_etal_2023a}, typically attributed to lower stellar and gas-phase metallicities, young stellar ages, and/or higher gas densities. Therefore, if we can constrain which of these dependencies primarily modulates $U$, we can use $U$ to directly trace how galaxies evolve. It is also essential to understand which dependency drives $U$ to reveal underlying details of the galaxies themselves. Of the proposed culprits for a redshift evolution of $U$, $n_{\rm{e}}$ is the easiest to directly measure and determine how it plays a role in modulating $U$ across cosmic time.

Electron densities can be directly measured from collisionally-excited transitions that share a common lower energy level, but are populated from a split fine structure upper energy level. This minuscule energy difference results in the emission of a doublet transition, and the relative excitation rates of each doublet line depend only on the collisional excitation rates with free electrons  \citep[in the low-density limit; e.g., ][]{osterbrock_1989, osterbrock_ferland_2006, draine_2011}, making the doublet ratio sensitive to the electron density. Examples of density-sensitive diagnostics include the ratios of the low-ionization rest-frame optical [\ion{O}{2}]$\lambda\lambda3727,3730 {\rm{\AA \AA}}$ and [\ion{S}{2}]$\lambda\lambda6718,6733 {\rm{\AA \AA}}$ ($E_{\rm{IP}}=13.61~\rm{and}~10.36$ eV, respectively) doublet lines \citep[e.g.][]{kewley_etal_2019}, which are typically strong in local star-forming galaxies and observable at $z>1$ after being redshifted into the rest-frame near-infrared (NIR). Ground-based NIR observations have allowed us to probe typical low-ionization electron densities using these tracers in multiple statistical samples of galaxies spanning $0<z<3$ \citep[e.g.][]{erb_etal_2006, liu_etal_2008}. From these studies, observational evidence suggests that galaxies became increasingly denser with increasing redshift at fixed stellar mass \citep[e.g.][]{steidel_etal_2014, sanders_etal_2016, kaasinen_etal_2017a, davies_etal_2021}, where $n_{\rm{e}}\sim200-300$ cm$^{-3}$ at $1<z<4$ as opposed to $n_{\rm{e}}\sim10-100$ cm$^{-3}$ at $z<1$. 

\begin{figure*}
\begin{center}
    \includegraphics[width=0.75\textwidth, trim=30 0 30 0,  clip=yes]{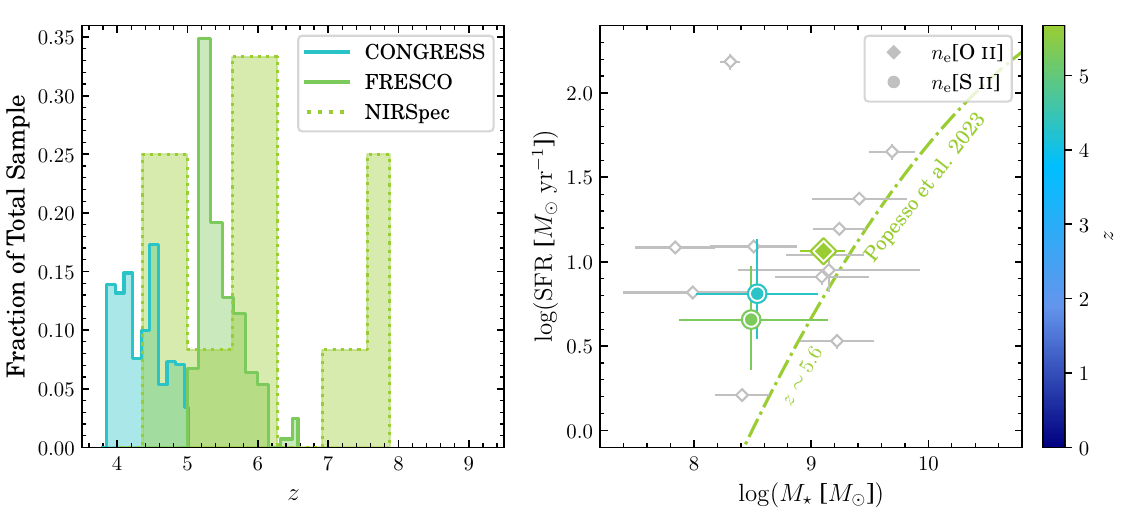}
    \caption{Sample statistics and galaxy demographics for the NIRSpec/MSA (light green), CONGRESS (teal), and FRESCO (dark green) datasets analyzed in this work. \textit{Left:} Histogram of the redshift distributions of each sample. \textit{Right:} Total stellar masses and star-formation rates, compared to the star-forming main sequence at similar redshifts \citep{popesso_etal_2023}, shown in green, for comparison. The individual galaxies in the NIRSpec/MSA sample are shown in light grey, and the sample medians are colored by median redshift. The three sample medians uniformly probe a redshift range from $4<z<8$ and sample galaxies that are slightly biased towards lower masses and higher star-formation rates at these redshifts.}
\label{fig:samp_props}
\end{center}
\end{figure*}

Prior to the launch of the \textit{JWST} \citep[][]{gardner_jwst_etal_2023}, detecting either doublet beyond $z\sim4$ for galaxy samples was difficult. High-$z$ galaxies have harder ionizing radiation fields that weaken low-ionization transitions, making these doublets intrinsically fainter \citep[e.g.][]{charlot_longhetti_2001}. This is compounded with the fact that many other key rest-frame optical emission lines that probe stellar and nebular properties (e.g. [\ion{O}{3}]$\lambda4363$\AA, [\ion{O}{3}]$\lambda5008$\AA, H$\alpha$) are redshifted to longer observed-frame wavelengths, and are therefore out of reach from the ground. \textit{JWST} has since enhanced our ability to understand the electron density and other nebular properties in $z>4$ galaxies \citep[e.g.][]{arellano-cordova_etal_2022, isobe_etal_2023, nakajima_etal_2023, reddy_etal_2023b, abdurrouf_etal_2024, hsiao_etal_2024a, hsiao_etal_2024b, sanders_etal_2024, yanagisawa_etal_2024, li_etal_2025, peng_etal_2025a, rogers_etal_2025, scholte_etal_2025, shapley_etal_2025a, stanton_etal_2025, topping_etal_2025}. Those focusing on [\ion{O}{2}] and [\ion{S}{2}] electron densities at $z>4$ have found higher densities than those at $z<4$, suggesting that electron density continues to increase with $z$ at high-$z$ \citep{isobe_etal_2023, reddy_etal_2023a, reddy_etal_2023b, abdurrouf_etal_2024, harikane_etal_2025b, topping_etal_2025, usui_etal_2025}. The increased $n_{\rm e}$ values have been suggested to arise due to higher overall gas densities, as well as metallicity, ionization structure, and geometrical effects. Higher gas densities could be driven by the higher average baryon density at higher redshift, scaling as $(1+z)^3$. By necessity, previous samples were small and focused on a bright subset of galaxies preselected with varying selection functions for NIRSpec follow-up. There has not yet been an unbiased and complete sample of $n_{\rm e}$ observed past $z\sim4$.

With \textit{JWST}, it is possible to disentangle the complex relationship between $n_{\rm{e}}$, $\epsilon$, $U$, and other galaxy properties, and how these influence the evolution of galaxies and their nebular conditions across cosmic time. This paper is organized as follows: In Section \ref{sec:nirspec_nircam}, we present new samples of NIRSpec/MSA and NIRCam/grism observations used to compute electron densities from $4<z<8$. We then compile a sample of literature $n_{\rm{e}}$ estimates to probe the evolution from $0<z<8$ in Section \ref{sec:lit}. In Section \ref{sec:methods}, we present the measurements of $n_{\rm{e}}$, $U$, and other galaxy properties. In Section \ref{sec:results} we discuss the evolution of $n_{\rm{e}}$ with $z$ and $U$, and the dependence of $n_{\rm{e}}$ on galaxy properties. Finally, in Section \ref{sec:discuss} we discuss the impact of our findings on the nebular structure of high-$z$ galaxies. We end by summarizing our main conclusions of this study in Section \ref{sec:fin}.

In this work, all magnitudes are AB magnitudes \citep{oke_gunn_1983}. Unless otherwise specified, all emission line wavelengths are reported as vacuum transitions, adopted from the Atomic Line List\footnote{\url{https://www.pa.uky.edu/~peter/newpage/}} \citep[][v3.00b5]{vanhoof_2018}, and emission line fluxes are reported in the rest-frame. We adopt the solar abundances of \citet{asplund_etal_2021} and assume a cosmology where $H_0=67.4~\rm{km}~s^{-1}~Mpc^{-1}$, $\Omega_{\rm m}=0.3$, and $\Omega_{\Lambda}=0.7$ \citep{planck_collab_2016}.

\begin{figure*}
\begin{center}
    \includegraphics[width=0.8\textwidth, trim=30 0 30 0,  clip=yes]{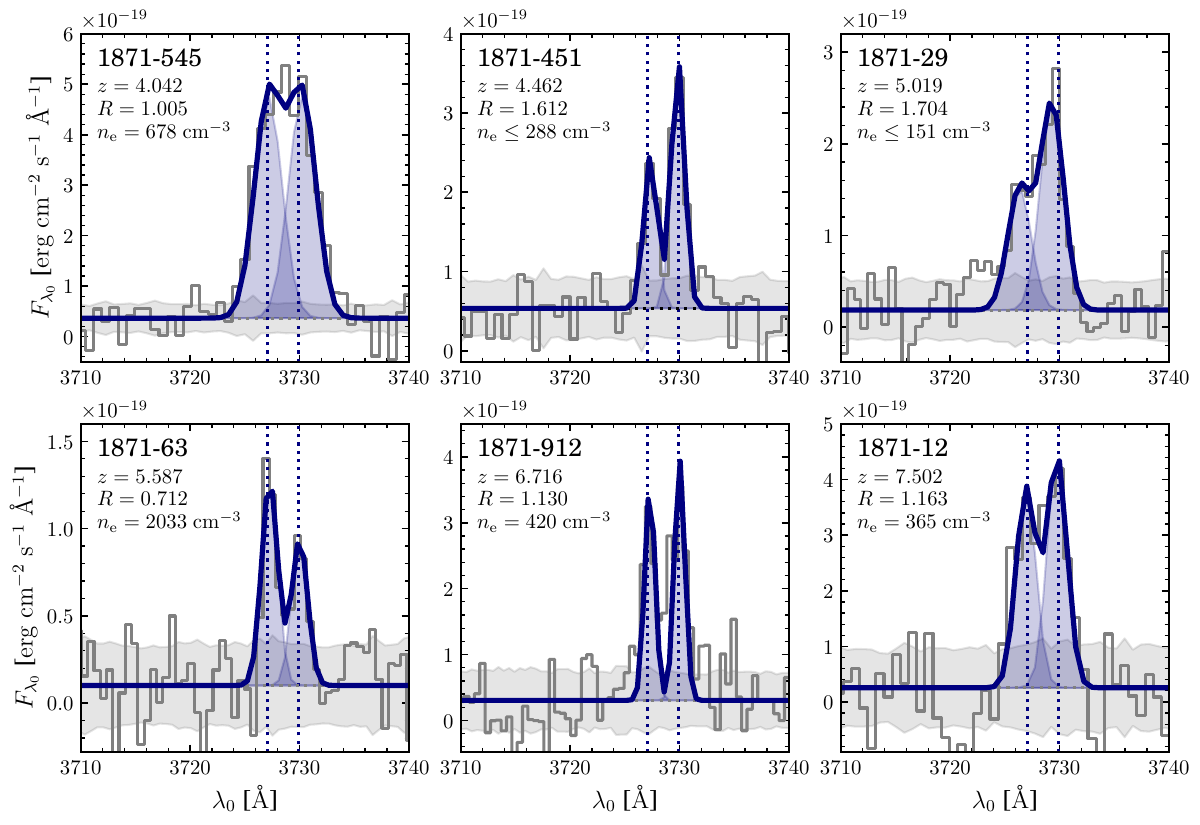}
        \caption{[\ion{O}{2}] $\lambda\lambda$3727, 3730\AA\AA~emission line profiles and Gaussian fits for the NIRSpec/MSA GO 1871 galaxies. In each panel, we provide the redshift of the galaxy ($z$), the measured [\ion{O}{2}] flux ratio ($R$), and the derived electron density ($n_{\rm{e}}$). The rest-frame wavelengths of the [\ion{O}{2}] $\lambda\lambda$3727, 3730\AA\AA~emission lines are shown by vertical dark blue dotted lines. The two doublet components for the Gaussian fits are shown as shaded dark blue, and the error on the data is shaded in grey.}
\label{fig:go_O2}
\end{center}
\end{figure*}

\section{The NIRSpec/MSA and NIRCam/Grism Samples} 
\label{sec:nirspec_nircam}

For the first time, the {\it{JWST}} \citep{mcelwain_etal_2023, rigby_etal_2023} has the spectroscopic resolution and depth to accurately constrain electron densities of galaxies at $z=4-8$. We curate these samples to exclusively contain galaxies with spectroscopically-resolved profiles of the rest-frame [\ion{O}{2}]$\lambda\lambda3727,3730$\AA\AA~and [\ion{S}{2}]$\lambda\lambda6718,6733$\AA\AA~doublets (with $\Delta v\sim224~\rm{km}~s^{-1}$ and $\Delta v\sim642~\rm{km}~s^{-1}$, respectively). Here, we prioritize datasets observed with instruments on {\it{JWST}} that meet the minimum line velocity separation to determine $n_{\rm e}$ at $4<z<8$, i.e. the NIRSpec H-gratings ($v_{\rm{res}}\sim111~\rm{km}~s^{-1}$) and the NIRCam grism ($v_{\rm{res}}\sim187~\rm{km}~s^{-1}$). In doing so, we are well-equipped to disentangle the underlying trends of galaxy properties with $n_{\rm e}$ at high-$z$. To provide a statistical sampling of $n_{\rm e}$ in these underexplored redshift regimes, we include three different samples observed with {\it{JWST}}, which we outline in the following subsections below. We also show the demographics of these samples in Figure \ref{fig:samp_props}, where we see that they uniformly probe redshifts between $4<z<8$ and consist of galaxies that have slightly lower masses and higher star-formation rates at these redshifts. We also note that we exclude galaxies from all these samples that exhibit signatures (i.e. broad lines) of active galactic nuceli, and only include star-forming galaxies.
\subsection{NIRSpec/MSA: Project ID GO 1871} 
\label{sec:go_1871}

We first summarize the details of the JWST Cycle 1 GO 1871 Program \citep[PI Chisholm;][]{chisholm_jwst_go1871, chisholm_etal_2024} here, which comprises observations with the {\it{JWST}}/NIRSpec Micro-Shutter Array (MSA). We refer the reader to the relevant sections in \citet{chisholm_etal_2024}, \citet{gazagnes_etal_2024}, and \citet{saldana-lopez_etal_2025} for more information and details regarding the program. While we focus on a subsample of galaxies observed by the program, the full GO 1871 sample is discussed in \citet{saldana-lopez_etal_2025}.

\subsubsection{Observations}
\label{sec:go_1871_obs}

The program targeted 20 galaxies in the GOODS-North (GN) field with photometric redshifts $z_{\rm{phot}}>5$ \citep{finkelstein_etal_2015, bouwens_etal_2015}, as well as some with spectroscopic redshifts measured from the Ly$\alpha$ transition \citep{jung_etal_2020}. They were also selected to be brighter than $m_{\rm{F160W}}\sim28$ mag in {\it{HST}} imaging and $>3\sigma$ detections in {\it{Spitzer}} 4.5 $\mu$m, which limits to galaxies that resembled typical star-forming galaxies that existed during the epoch of reionization. The targets were observed with the \textit{JWST}/NIRSpec \citep{böker_etal_2023} filter-grating pairs F170LP-G235H and F290LP-G395H using the MSA \citep{ferruit_etal_2022} on February 10, 2023. We used the high-resolution ($R\sim2700$, $v_{\rm{res}}\sim111~\rm{km}~s^{-1}$) gratings to detect velocity-resolved profiles of the rest-frame \ion{Mg}{2} emission lines \citep{henry_etal_2018, chisholm_etal_2020}. Observations with the MSA were centered on the galaxy 1871-12 \citep{finkelstein_etal_2013, hutchison_etal_2019, jung_etal_2020}, a distinct Ly$\alpha$ emitter at $z=7.5$ with a robust detection of the \ion{Mg}{2} doublet, presented and discussed further in \citet{gazagnes_etal_2024}.

\begin{deluxetable*}{cccccccccccc}
\tablecaption{NIRSpec Density Sample: Observed and Derived Properties}
\tabletypesize{\footnotesize}
\tablewidth{0pt}
\tablehead{
\\[-1.0pc]
\colhead{Galaxy ID} & \colhead{RA} & \colhead{Dec} & \colhead{$z$} & \colhead{$R$[\ion{O}{2}]} & \colhead{$n_{\rm{e}}$[\ion{O}{2}]} & \colhead{$\log(M_{\star})$} & \colhead{$\log({\rm{SFR}})$} & \colhead{$\log({\rm{sSFR}})$} & \colhead{$\log({\Sigma_{\rm{SFR}}})$} & \colhead{$\log({\rm{O}}_{32})$}
\\[-0.6pc]
\colhead{} & \colhead{[deg]} & \colhead{[deg]} & \colhead{} & \colhead{} & \colhead{[cm$^{-3}$]} & \colhead{[$M_{\odot}$]} & \colhead{[$M_{\odot}~{\rm{yr}}^{-1}$]} & \colhead{[${\rm{Gyr}}^{-1}$]} & \colhead{[$M_{\odot}~{\rm{yr}}^{-1}~{\rm{kpc}}^{-2}$]} & \colhead{}
}
\startdata
1871-12 & 189.15797785 & 62.30239809 & $7.5020$ & $1.163\pm0.289$ & $365^{+163}_{-80}$ & $8.31^{+0.07}_{-0.05}$ & $2.19^{+0.04}_{-0.05}$ & $2.87\pm 0.10$ & $1.31\pm 0.04$ & $0.76\pm0.43$ \\
1871-29 & 189.18882760 & 62.26991420 & 5.0191 & $1.704\pm0.378$ & $\leq151$ & $9.24^{+0.12}_{-0.19}$ & $1.19^{+0.01}_{-0.01}$ & $0.95\pm 0.23$ & $-0.15\pm 0.06$ & $0.68\pm0.47$ \\
1871-63 & 189.17219338 & 62.30563914 & 5.5872 & $0.712\pm0.220$ & $2033^{+830}_{-275}$ & $7.84^{+0.32}_{-0.12}$ & $1.08^{+0.02}_{-0.02}$ & $2.24\pm 0.34$ & $\geq0.96$ & $1.48\pm1.34$ \\
1871-451 & 189.23835414 & 62.28442312 & 4.4615 & $1.612\pm0.397$ & $\leq288$ & $7.99^{+0.40}_{-0.44}$ & $0.82^{+0.02}_{-0.03}$ & $1.82\pm 0.60$ & $-0.10\pm 0.05$ & $0.34\pm0.50$ \\
1871-545 & 189.14071083 & 62.27725108 & 4.0416 & $1.005\pm0.076$ & $678^{+56}_{-40}$ & $9.41^{+0.25}_{-0.32}$ & $1.37^{+0.01}_{-0.01}$ & $0.96\pm 0.41$ & $0.08\pm 0.03$ & $-0.14\pm0.11$ \\
1871-912 & 189.18613720 & 62.27089253 & 6.7161 & $1.130\pm0.205$ & $420^{+124}_{-72}$ & $9.69^{+0.10}_{-0.17}$ & $1.65^{+0.05}_{-0.05}$ & $0.96\pm 0.21$ & $0.0\pm 0.04$ & $0.82\pm0.35$ \\
GLASS-10021 & 3.608511 & --30.418541 & 7.2863 & $1.400\pm 0.186$ & $76^{+196}_{-76}$ & $8.51^{+0.30}_{-0.22}$ & $1.09^{+0.02}_{-0.02}$ & $1.58\pm 0.37$ & -- & $1.08\pm0.36$ \\
GLASS-50038 & 3.565199 & --30.394264 & 5.7720 & $1.686\pm 0.253$ & $\leq45$ & $9.09^{+0.39}_{-0.09}$ & $0.91^{+0.04}_{-0.05}$ & $0.82\pm 0.41$ & -- & $0.59\pm0.66$ \\
GLASS-100001 & 3.603845 & --30.382234 & 7.8732 & $1.461\pm 0.227$ & $21^{+216}_{-21}$ & $9.15^{+0.55}_{-0.55}$ & $0.95^{+0.10}_{-0.13}$ & $0.80\pm 0.79$ & -- & $0.45\pm0.39$ \\
GLASS-110000 & 3.570642 & --30.414638 & 5.7641 & $0.645\pm 0.111$ & $2747^{+2437}_{-1029}$ & $9.22^{+0.16}_{-0.27}$ & $0.53^{+0.04}_{-0.05}$ & $0.31\pm 0.32$ & -- & $0.81\pm0.59$ \\
GLASS-150029 & 3.577166 & --30.422576 & 4.5838 & $1.417\pm 0.145$ & $60^{+138}_{-60}$ & $9.12^{+0.03}_{-0.33}$ & $1.04^{+0.02}_{-0.02}$ & $0.92\pm 0.33$ & -- & $1.08\pm0.40$ \\
GLASS-160122 & 3.564901 & --30.424956 & 5.3319 & $1.091\pm 0.370$ & $492^{+1335}_{-435}$ & $8.41^{+0.16}_{-0.17}$ & $0.21^{+0.02}_{-0.01}$ & $0.80\pm 0.23$ & -- & -- \\[0.25pc]
\hline \\[-0.6pc]
Median &  &  & $5.676^{+2.197}_{-1.634}$ & $1.282^{+0.096}_{-0.106}$ & $202^{+145}_{-102}$ & $9.11^{+0.17}_{-0.19}$ & $1.06^{+0.02}_{-0.02}$ & $0.96^{+0.14}_{-0.13}$ & $0.04^{+1.17}_{-1.10}$ & $0.76^{+0.30}_{-0.34}$ \\[0.25pc]
\enddata
\tablecomments{RA and Dec are the right ascension and declination in J2000 coordinates, respectively. The bottom row reports the medians of each parameter. Median redshift error bars indicate redshift spread of sample, not the error on the median.}
\label{tab:nirspec_tab}
\end{deluxetable*}

The observed wavelength coverage of the
F170LP-G235H and F290LP-G395H filter-grating pairs ($1.66-3.05~\mu$m and $2.87-5.14~\mu$m, respectively) provides access to the rest-frame transitions of [\ion{O}{2}]$\lambda\lambda3727,3730$\AA\AA~and/or [\ion{S}{2}]$\lambda\lambda6718,6733$\AA\AA~for objects at $z\sim4-8$. Low-ionization emission lines are observed to be intrinsically weak at high-$z$ due to the hard radiation fields of metal-poor stars \citep[e.g.][]{erb_etal_2006, strom_etal_2017}. The NIRSpec observations were divided between the two selected filter-grating pairs, with more exposure time allocated to the G235H grating, 53,044 seconds (or approximately 14.7 hours) in 36 exposures, in order to detect the faint \ion{Mg}{2}~doublet \citet{gazagnes_etal_2024}. The G395H grating exposure times are significantly shorter than those with G235H, 9716 seconds (or approximately 2.7 hours) in 6 exposures, which was expected to cover rest-frame optical emission lines, including the [\ion{O}{2}] and [\ion{S}{2}] doublets. All exposures were read out using the NRSIRS2 readout mode.

\subsubsection{Data Reduction}
\label{sec:go_1871_specred}

We reduced the NIRSpec/MSA data products with the Python data reduction package {\texttt{msaexp}}\footnote{\url{https://github.com/gbrammer/msaexp}} \citep[][v.0.8.4]{brammer_etal_2022}, which is a wrapper for the NIRSpec reduction pipeline. We implemented the reference files listed under \texttt{jwst\_1235.pmap}, which were made available on the CDRS website in May 2024. The {\texttt{msaexp}} software applies 1/f noise corrections, removes biases using a median, and re-scales noise based on empty regions of the exposures. It then runs several Level 2 {\it{JWST}} calibration pipeline functions, using the standard Space Telescope Science Institute data reduction pipeline version 1.14.0, and uses the local background from the nodded slits. To detect a sufficient background for subtraction, the observations employ the standard three-shutter nodding pattern. The final data co-additions for each target include all exposures drizzled onto a common grid. We compared the noise array with a running standard deviation of the flux using {\texttt{msaexp}} and found good agreement \citep{gazagnes_etal_2024}. We also apply the default point-source slitloss correction for all sources, and check their consistency with the photometry using \texttt{msafit}\footnote{\url{https://github.com/annadeg/jwst-msafit}} \citep{degraaff_etal_2024}. All emission lines are extracted using consistent optimal extractions.

\subsubsection{Density Sample Selection}
\label{sec:go_1871_samp}

Once we have reduced the NIRSpec data, we select galaxies for the final GO 1871 sample. The spectroscopic redshift of each galaxy was measured as the median redshift of the H$\alpha$, [\ion{O}{3}]$\lambda4960$\AA, [\ion{O}{3}]$\lambda5008$\AA, and [\ion{Ne}{3}]$\lambda3970$\AA~emission lines. Next, we apply a signal-to-noise ratio (S/N) cut to the sample, where all galaxies with [\ion{O}{2}] and [\ion{S}{2}] must have S/N $\geq3$ detections of each individual component in the doublet in order to be included in the analysis (see \S\ref{sec:meas} for details on emission line flux and uncertainty measurements). We utilize the Atomic Line List \citep[][v3.00b5]{vanhoof_2018} to verify the vacuum wavelengths of the ionic transitions. Applying these cuts gives us six galaxies, two of which are 1871-63 \citep[GN 42437;][]{chisholm_etal_2024} and 1871-12 \citep[GN 42912;][]{gazagnes_etal_2024}. We present the final sample of selected galaxies from GO 1871 with detections of the [\ion{O}{2}] doublet in Figure \ref{fig:go_O2}. These galaxies, along with the GLASS subsample containing six additional galaxies presented in \S\ref{sec:highz}, are subsequently referred to as the ``NIRSpec/MSA Sample" for the remainder of the paper. We also detected both components of the [\ion{S}{2}] doublet in the galaxy 1871-545, but found no evidence for [\ion{S}{2}] detections in any other galaxy. This is the only galaxy in the sample, and one of the highest redshift sources in the literature ($z=4.042)$, with detections of both [\ion{O}{2}] and [\ion{S}{2}]. Each NIRSpec/MSA galaxy and their observed and derived parameters are reported in Table \ref{tab:nirspec_tab}.

\subsection{NIRCam/Grism: CONGRESS \& FRESCO}
\label{sec:congress_fresco}

To supplement the NIRSpec/MSA sample, we include two slitless grism surveys using the \textit{JWST}/NIRCam grism \citep{rieke_etal_2005, rieke_etal_2023}. These are the Complete NIRCam Grism Redshift Survey \citep[CONGRESS; Cycle 2 GO 3577; PI Egami;][]{egami_jwst_go3577}, and the First Reionization Epoch Spectroscopically Complete Observations Survey \citep[FRESCO; Cycle 1 GO 1895; PI Oesch;][]{oesch_etal_2023}, which are blind, emission line luminosity-complete surveys  \citep{oesch_etal_2023}. FRESCO obtained NIRCam imaging and slitless grism spectroscopy in the GOODS-North and GOODS-South fields, and CONGRESS covered the same regions as FRESCO in GOODS-North \citep{covelo-paz_etal_2025}. For this study we only include NIRCam/grism spectra taken in GOODS-North (GN) field. For more information regarding these surveys, we refer the reader to the FRESCO survey paper by \citet{oesch_etal_2023} and the H$\alpha$ emitters from \citet{covelo-paz_etal_2025}.

\subsubsection{Observations}
\label{sec:congress_fresco_obs}

Both surveys observed the GN field, using the F356W filter for CONGRESS \citep{egami_jwst_go3577}, and the F444W filter \citep{oesch_etal_2023} for FRESCO. This provides a combined wavelength coverage ($3.1-5.0~\mu$m), enabling a blind search for reionization-era H$\alpha$ emitters at $3.7<z<6.7$ \citep{covelo-paz_etal_2025}. The CONGRESS sources were observed on January 9, 2024 with 8 pointings for 3779 seconds each (or approximately 1.0 hours; Sun et al. in preparation). The FRESCO sources were observed on February 7, 2023 with 8 pointings, each for 7043 seconds  \citep[or approximately 2.0 hours;][]{oesch_etal_2023}. \citet{covelo-paz_etal_2025} determined that both CONGRESS and FRESCO are luminosity-complete down to log$(L_{\rm H\alpha}/\rm{erg}~s^{-1})\sim41.75$, corresponding to ${\rm{SFR}}\sim3~M_\odot/ \rm yr^{-1}$. This ensures that we are observing all galaxies above this detection threshold with moderate spectral resolution ($R\sim1600$, $v_{\rm{res}}\sim187~\rm{km}~s^{-1}$). CONGRESS and FRESCO provide us with immense power in probing the electron density using a large and complete sample of galaxies at $z>4$.

\subsubsection{Data Reduction}
\label{congress_fresco_specred}

The NIRCam/grism data products were reduced using \texttt{grizli}\footnote{\url{https://github.com/gbrammer/grizli}} \citep{brammer_etal_2018}, using the \texttt{jwst\_1123.pmap} reference files, made available on the CDRS website in September 2023, as well as with modified sensitivity functions from \texttt{grizli} \citep{covelo-paz_etal_2025}. We also implement a two-step median filtering routine to continuum-subtract the data and isolate the emission lines as described in \citet{kashino_etal_2023}, \citet{meyer_etal_2024}, and \citet{kotiwale_etal_2026}. The first step is to identify detected emission lines (defined as pixels with S/N $>$ 3) in order to mask them for the final continuum subtraction, and is performed with a kernel (71 pixels) along the dispersion axis that has a hole in the center (10 pixels wide). Then a second median filter subtraction is run with these pixels flagged, improving the continuum subtraction around emission lines. Since the redshifted H$\alpha$ line is the only strong line at $3.7<z<6.7$ in the observed wavelength range, we have a three step process to select potential galaxies: 1) sources with photometric redshifts determined from \texttt{EAZY}\footnote{\url{https://github.com/gbrammer/eazy-photoz}} \citep{brammer_etal_2008} at $z_{\rm{phot}}=3.7-5.2$ for the F365W CONGRESS field and $z_{\rm{phot}}=4.85-6.7$ for the F444W FRESCO field, with redshift uncertainties less than 20\%, 2) sources with $m_{\rm F356W}~{\rm{and}}~m_{\rm F444W} <28$ mag, and 3) extract all sources with ancillary spectroscopic redshifts with the same ranges as above. \citet{covelo-paz_etal_2025} then used \texttt{grizli} to search for emission lines with $0.02(1+z_{\rm{spec}})$ of the \texttt{EAZY} redshift.

\subsubsection{Density Sample Selection}
\label{sec:congress_fresco_samp}

\begin{figure*}
\begin{center}
    \includegraphics[width=0.8\textwidth, trim=30 0 30 0, clip=yes]{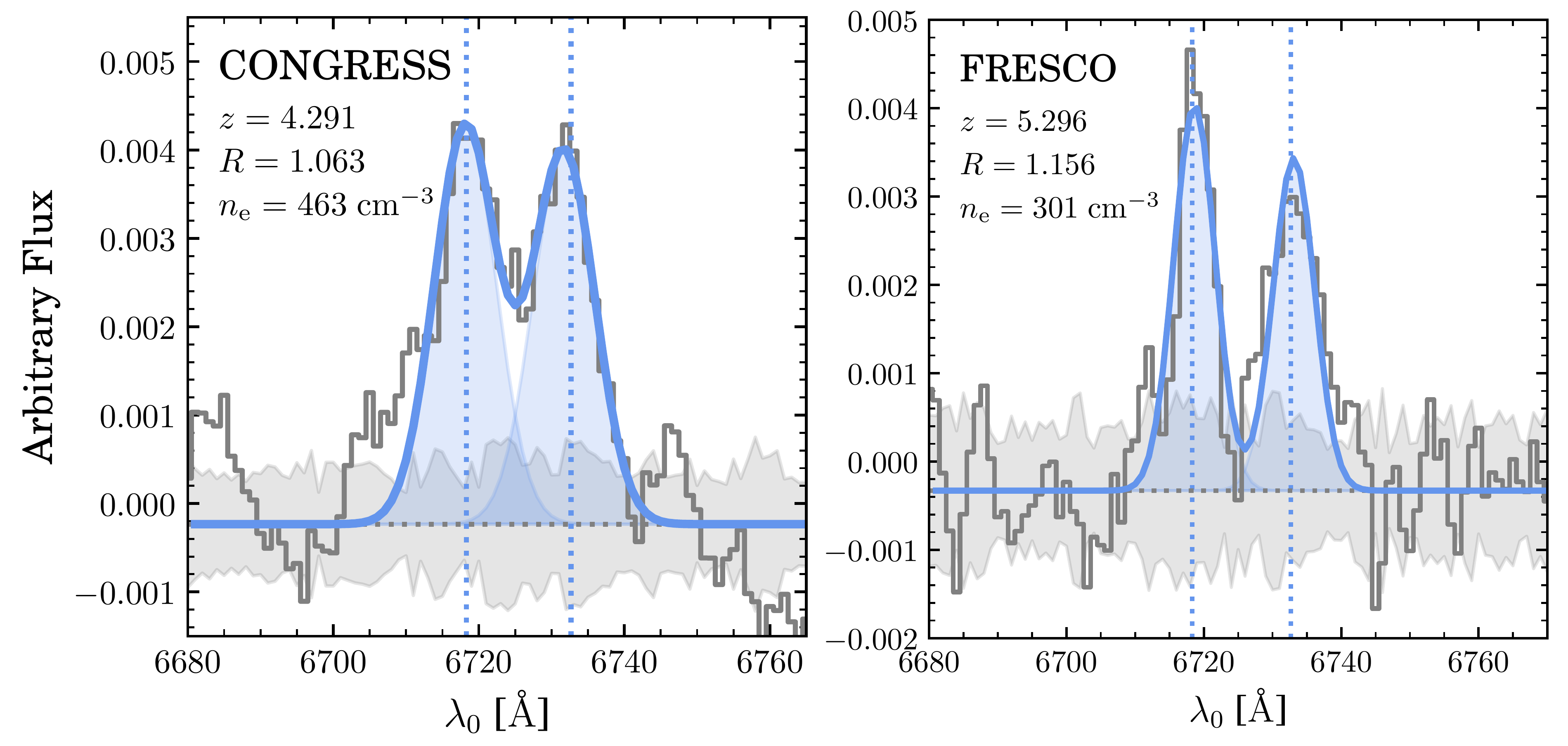}
        \caption{Stacked [\ion{S}{2}]$\lambda\lambda$6718, 6733\AA\AA~emission line profiles and Gaussian fits for the NIRCam/grism CONGRESS (\textit{left}) and FRESCO (\textit{right}) samples, respectively. In each panel, we provide the median redshift ($z$) of the stacks, the measured [\ion{S}{2}] flux ratio ($R$), and the derived electron density ($n_{\rm{e}}$). The rest-frame wavelengths of the [\ion{S}{2}]$\lambda\lambda$6718, 6733\AA\AA~emission lines are shown by vertical light blue dotted lines. The two doublet components for the Gaussian fits are shown as shaded light blue, and the error on the data is shaded in grey.}
\label{fig:congress_fresco_S2}
\end{center}
\end{figure*}

\begin{deluxetable*}{ccccccccc}
\tablecaption{NIRCam Density Sample: Observed and Derived Properties}
\tabletypesize{\footnotesize}
\tablewidth{0pt}
\tablehead{
\\[-1.2pc]
\colhead{Survey} & \colhead{$N$} & \colhead{$z$} & \colhead{$R$[\ion{S}{2}]} & \colhead{$n_{\rm{e}}$[\ion{S}{2}]} & \colhead{$\log({M_{\star}})$} & \colhead{$\log({\rm{SFR}})$} & \colhead{$\log({\rm{sSFR}})$} & \colhead{$\log(\Sigma_{\rm{SFR}})$}
\\[-0.6pc]
\colhead{} & \colhead{} & \colhead{} & \colhead{} & \colhead{[cm$^{-3}$]} & \colhead{[$M_{\odot}$]} & \colhead{[$M_{\odot}~{\rm{yr}}^{-1}$]} & \colhead{[${\rm{Gyr}}^{-1}$]} & \colhead{[$M_{\odot}~{\rm{yr}}^{-1}~{\rm{kpc}}^{-2}$]}
}
\startdata
CONGRESS & 410 & $4.291 ^{+0.723}_{-0.500}$ & $1.063\pm0.120$ & $463^{+281}_{-203}$ & $8.48^{+0.65}_{-0.61}$ & $0.65^{+0.32}_{-0.30}$ & $1.17^{+0.58}_{-0.53}$ & $-1.33^{+0.08}_{-0.02}$ \\
FRESCO & 281 & $5.296^{+1.276}_{-0.385}$ & $1.156\pm0.186$ & $301^{+388}_{-206}$ & $8.54^{+0.52}_{-0.52}$ & $0.81^{+0.33}_{-0.27}$ & $1.27^{+0.52}_{-0.43}$ & $-1.59^{+0.11}_{-0.03}$ \\[0.25pc]
\enddata
\tablecomments{The sample statistics of the stacked data. The columns give number ($N$), the median redshift, the [\ion{S}{2}] intensity ratio ($R[$\ion{S}{2}]), and the inferred electron density ($n_{\rm e}$). The last four columns give the median host galaxy properties of each sample. Median redshift error bars indicate redshift spread of sample, not the error on the median.}
\label{tab:nircam_tab}
\end{deluxetable*}

To select preliminary sources after performing the line search, \citet{covelo-paz_etal_2025} apply a S/N threshold, where all sources with an integrated H$\alpha$ S/N $\leq5$ are not included. After accounting for galaxies with multiple components, recomputing photometry, and performing a new extraction with \texttt{grizli}, each source was then visually inspected by four CONGRESS and FRESCO team members using \texttt{specvizitor}\footnote{\url{https://github.com/ivkram/specvizitor}}. Following the classification presented in \citet{meyer_etal_2024}, each source was assigned a quality flag by the individual team members ranging from $q=3$ to $q=-1$ \citep{covelo-paz_etal_2025}, and then given a final quality flag as the average of the four individual scores. We selected sources with an overall quality flag score of $q\geq1.75$ for the final H$\alpha$ catalog, indicating that a majority of the team members believed that the observed H$\alpha$ lines were likely real detections. This results in a catalog of 532 H$\alpha$ emitters in CONGRESS and 305 in FRESCO in the GN field \citep{covelo-paz_etal_2025}. 

To create our final electron density sample, we start with these sources. We stack the spectra of all the CONGRESS and FRESCO sources, respectively, to obtain the median properties of each sample. In addition, the [\ion{S}{2}] doublet is not detected in all individual CONGRESS and FRESCO sources, so by stacking the spectra we are able to measure a median [\ion{S}{2}] line ratio. To do this, we normalize each continuum-subtracted spectrum by the integrated H$\alpha$ luminosity, which reduces the bias of bright galaxies that contribute to the stacked spectra. Next, we convert all wavelengths, flux densities and errors to rest-frame using our measured H$\alpha$ wavelength. We then perform a median stack to create a single spectrum for both CONGRESS and FRESCO and bootstrap the median uncertainties 10,000 times. In the final CONGRESS and FRESCO stacked spectra we end up with a total of 410 and 281 sources, respectively. These stacks, shown in Figure \ref{fig:congress_fresco_S2}, create a large luminosity-complete sample to compute the median electron densities of galaxies at $z>4$. Both NIRCam grism samples and their observed and derived parameters are reported in Table \ref{tab:nircam_tab}. Both the CONGRESS and FRESCO stacked spectra only cover the [\ion{S}{2}] doublet and not the [\ion{O}{2}] doublet.

\begin{deluxetable*}{cccccccccccc}[]
\rotate
\tablecaption{Density Samples from the Literature: Observed and Derived Properties}
\tabletypesize{\footnotesize}
\tablewidth{0pt}
\tablehead{
\\[-1.0pc]
\colhead{Survey} & \colhead{$N$} & \colhead{$z$} & \colhead{$R$[\ion{O}{2}]} & \colhead{$n_{\rm{e}}$[\ion{O}{2}]} & \colhead{$R$[\ion{S}{2}]} & \colhead{$n_{\rm{e}}$[\ion{S}{2}]} & \colhead{$\log({M_{\star}})$} & \colhead{$\log({\rm{SFR}})$} & \colhead{$\log({\rm{sSFR}})$} & \colhead{$\log(\Sigma_{\rm{SFR}})$} & \colhead{$\log({\rm{O}}_{32})$}
\\[-0.6pc]
\colhead{} & \colhead{} & \colhead{} & \colhead{} & \colhead{[cm$^{-3}$]} & \colhead{} & \colhead{[cm$^{-3}$]} & \colhead{[$M_{\odot}$]} & \colhead{[$M_{\odot}~{\rm{yr}}^{-1}$]} & \colhead{[${\rm{Gyr}}^{-1}$]} & \colhead{[$M_{\odot}~{\rm{yr}}^{-1}~{\rm{kpc}}^{-2}$]} & \colhead{}
}
\startdata
SDSS DR7 & 110,106 & $0.0820^{+0.2868}_{-0.0820}$ & -- & -- & $1.406$ & $15^{+23}_{-15}$ & $10.16^{+0.04}_{-0.04}$ & $0.09^{+0.00}_{-0.00}$ & $-0.98^{+0.00}_{-0.00}$ & -- & $-0.77\pm 0.00$ \\
SDSS DR16 & 32,019 & $0.1434^{+0.4218}_{-0.1432}$ & -- & -- & $1.381$ & $37^{+57}_{-37}$ & $10.34^{+0.07}_{-0.06}$ & $-0.08^{+0.00}_{-0.00}$ & $-1.42^{+0.03}_{-0.03}$ & -- & $-0.46\pm 0.00$ \\
LzLCS & 36 & $0.3105^{+0.0734}_{-0.0912}$ & -- & -- & $1.335^{+0.070}_{-0.073}$ & $81^{+92}_{-77}$ & $9.12^{+0.14}_{-0.15}$ & $1.26^{+0.00}_{-0.00}$ & $1.14^{+0.04}_{-0.04}$ & $0.96^{+0.02}_{-0.02}$ & $0.43\pm 0.12$ \\
CLASSY & 45 & $0.022^{+0.160}_{-0.020}$ & -- & -- & $1.300^{+0.010}_{-0.011}$ & $119^{+41}_{-41}$ & $8.13^{+0.29}_{-0.26}$ & $0.17^{+0.00}_{-0.00}$ & $1.04^{+0.04}_{-0.04}$ & $-0.87$ & $0.32\pm 0.13$ \\
MOSDEF & 69* & $2.24^{+1.50}_{-0.92}$ & $1.18^{+0.01}_{-0.10}$ & $339^{+175}_{-15}$ & $1.13^{+0.16}_{-0.06}$ & $343^{+105}_{-200}$ & $10.0$ & $1.38$ & $0.38$ & -- & $0.10\pm 0.37$ \\
COSMOS-[\ion{O}{2}] & 21 & $1.524^{+0.130}_{-0.118}$ & $1.29^{+0.03}_{-0.03}$ & $191^{+37}_{-37}$ & -- & -- & $10.59$ & $1.45$ & $-0.14$ & -- & -- \\
KBSS-MOSFIRE & 171 & $2.35^{+0.4}_{-0.4}$ & $1.13^{+0.03}_{-0.03}$ & $420^{+54}_{-53}$ & -- & -- & $9.98$ & $1.36$ & $0.38$ & -- & -- \\
SAMI \& KMOS$^{3 \rm D}$ & 471 & $\leq0.1$ & -- & -- & $1.40^{+0.01}_{-0.01}$ & $21^{+16}_{-16}$ & -- & -- & -- & -- & -- \\
& 39 & $0.6<z<1.1$ & -- & -- & $1.32^{+0.11}_{-0.08}$ & $95^{+88}_{-78}$ & -- & -- & -- & -- & -- \\
& 36 & $1.1<z<1.9$ & -- & -- & $1.34^{+0.05}_{-0.13}$ & $76^{+142}_{-53}$ & -- & -- & -- & -- & -- \\
& 65 & $1.9<z<2.6$ & -- & -- & $1.22^{+0.15}_{-0.10}$ & $211^{+143}_{-145}$ & -- & -- & -- & -- & -- \\
AURORA & 31 & $2.1023^{+0.3811}_{-0.7190}$ & -- & -- & $1.210^{+0.027}_{-0.027}$ & $224^{+38}_{-34}$ & $9.26^{+0.06}_{-0.05}$ & $1.12^{+0.09}_{-0.12}$ & $0.86^{+0.07}_{-0.09}$ & $-0.14$ & $0.18$ \\
& 16 & $3.1338^{+0.3251}_{-0.2119}$ & -- & -- & $1.105^{+0.063}_{-0.060}$ & $387^{+118}_{-103}$ & $8.99^{+0.072}_{-0.062}$ & $0.79^{+0.08}_{-0.08}$ & $0.80^{+0.08}_{-0.08}$ & $0.16$ & $0.31$ \\
& 4 & $5.3541^{+0.4686}_{-0.9426}$ & -- & -- & $1.005^{+0.187}_{-0.166}$ & $593^{+532}_{-324}$ & $8.82^{+0.07}_{-0.18}$ & $1.58^{+0.06}_{-1.51}$ & $1.76^{+0.07}_{-0.14}$ & $1.18$ & $0.65$ \\[0.25pc]
    \enddata
\tablecomments{Here, we report in order in which each sample is discussed in the text of \S\ref{sec:lit}. Median redshift error bars indicate redshift spread of sample, not the error on the median. *Sum of all galaxies used in the sample, please refer to the text for the number of galaxies used to derive each median density}
\label{tab:lit_tab}
\end{deluxetable*}

\section{Literature Samples}
\label{sec:lit}

To analyze trends of electron density with other nebular properties of galaxies across redshifts $0<z<8$, and to better contextualize our current understanding of early star and galaxy formation, we include several samples of galaxies at $z<4$ from the literature that have robust and resolved detections of [\ion{O}{2}]$\lambda\lambda$3727,3730\AA\AA~and [\ion{S}{2}]$\lambda\lambda$6718,6733\AA\AA. By ensuring that each sample 1) contains objects that have spectrally resolved emission line doublets, and 2) have emission line measurements or ratios using a consistent and homogeneous methodology, we have a collection of galaxy samples that cover a large range of redshifts with robust measurements of the electron density. In this section we will discuss each sample in detail, organized roughly by increasing median redshift. Since most of the literature datasets report the associated median values of the diagnostic line ratio, we adopt the median [\ion{O}{2}] and [\ion{S}{2}] line ratios as a typical value at a given redshift range, wherever applicable for each sample. For samples with galaxies that have individual [\ion{O}{2}] or [\ion{S}{2}] line fluxes (\S\ref{sec:local} and \S\ref{sec:highz}), we first compute the median line ratios for each galaxy, then compute the $n_{\rm e}$ from that median line ratio. Table \ref{tab:lit_tab} lists all literature sample properties used in this work.

\subsection{Local Samples}
\label{sec:local}

We begin by discussing the local ($z\sim0$) galaxy samples we use to derive electron densities of galaxies and examine how they play a role in galaxy evolution.

We compile our first two local literature samples from the Sloan Digital Sky Survey \citep[SDSS;][]{york_etal_2000} Data Release 7 \citep[DR7;][]{abazajian_etal_2009} and Data Release 16 \citep[DR16;][]{ahumada_etal_2020}. The redshifts, emission line measurements and total stellar masses of the SDSS DR7 sample are taken from the MPA-JHU catalogs \citep{kauffmann_etal_2003, brinchmann_etal_2004, tremonti_etal_2004}, and from the Portsmouth catalogs for the SDSS DR16 sample \citep{maraston_etal_2013, thomas_etal_2013}. At the resolution of SDSS, the [\ion{O}{2}] doublet is not resolved, so we make use of the [\ion{S}{2}] doublet line fluxes measured from both SDSS DR7 and DR16. To gather sufficient and representative samples, we include all star-forming galaxies with $z\geq0$. We also make cuts to the samples based on the S/N of [\ion{S}{2}] $\lambda\lambda$6718,6733\AA\AA, and galaxies with S/N $\geq5$ detections of each individual component in the doublet are included in the sample. Finally, we remove any duplicate objects from the parent SDSS samples that overlap with LzLCS and CLASSY (detailed below). As a result, the SDSS DR7 and DR16 samples have a total of 110,106 and 32,019 sources at median redshifts of $z=0.0819$ and $z=0.1434$, respectively. These two samples provide robust statistical anchors for typical local galaxy properties. 

% Recent JWST observations have shown that a large fraction of high-redshift galaxies show signs of hosting an active galactic nucleus \citep[AGN; e.g.][]{akins_etal_2024, matthee_etal_2024, kocevski_etal_2024, maiolino_etal_2024}, and there exist galaxies included in the NIRSpec/MSA and NIRCam/grism samples reported to house an AGN \citep[e.g.][]{chisholm_etal_2024}. Therefore, we choose to not separate AGN or starbursts, as this collection of SDSS galaxies is intended to serve as a ``global" sample of galaxies at $z\sim0$. \citet{flury_moran_2020} found that AGN in the SDSS have [\ion{S}{2}] densities systematically higher by $\sim0.5$ dex than their star-forming galaxy counterparts.
We also make use of other local galaxy samples compiled from the SDSS. The first is the Low-Redshift Lyman Continuum Survey archive sample \citep[LzLCS+;][]{flury_etal_2022, izotov_etal_2016a, izotov_etal_2016b, izotov_etal_2018a, izotov_etal_2018b, izotov_etal_2021, wang_etal_2019}. At $z\sim0.3$, the full LzLCS+ sample is comprised of 89 star-forming galaxies and aims to probe the escape of rest-frame Lyman continuum (LyC, $\lambda\leq912$\AA) photons, as well as the driving mechanisms and resulting effects. The sample is primarily made of low-mass galaxies, and samples extreme galaxy properties, such as high O$_{32}$. Despite this, LzLCS+ samples a wide range of stellar masses, $10^{8.25}\leq M_{\star} \leq 10^{10.75}$. For more information on the selection and completeness of the LzLCS+ sample, we refer the reader to the relevant sections in \citet{flury_etal_2022}. Since the optical emission line fluxes are pulled from SDSS, we make use of the [\ion{S}{2}] doublet line fluxes to measure a median electron density. We apply a S/N cut to the same emission lines as we did for the SDSS sample, where all galaxies that do not have lines with S/N $\geq5$ are not included in the final sample. After applying these cuts, the final LzLCS+ density sample contains 36 sources at a median redshift of $z=0.3105$. 

We also include the Cosmic Origins Spectrograph (COS) Legacy Archive Spectroscopic SurveY \citep[CLASSY;][]{berg_etal_2022, james_etal_2022} in our selection of local galaxy samples. Similarly with LzLCS+, it also contains local galaxies to serve as high-$z$ analogs. At $z\sim0$, the sample is comprised of 45 star-forming galaxies and aims to provide the first high-quality, high-resolution spectral database in the far-ultraviolet (UV) for studies of star-forming galaxies across cosmic time. The sample also primarily comprises galaxies with lower stellar masses, but still spans a wide range, $10^{6.2}M_\odot\leq M_{\star} \leq 10^{10.1}M_\odot$. For a more detailed discussion on the sample selection and completeness of the CLASSY sample, we refer the reader to the relevant sections in \citet{berg_etal_2022}, as well as subsequent sample papers \citep[e.g.][]{mingozzi_etal_2022, arellano-cordova_etal_2025}. Since the optical emission line fluxes are also pulled from SDSS, we make use of the [\ion{S}{2}] doublet line fluxes to measure a median electron density for the sample. Similar to the SDSS and LzLCS+ samples, we apply the same S/N cut, where all galaxies that do not have lines with S/N $\geq5$ are not included in the final sample. All CLASSY galaxies meet these requirements, so our final CLASSY density sample includes all 45 galaxies at a median redshift of $z=0.022$. We do note that a few of the CLASSY galaxies have optical spectra do not have SDSS spectra, and suggest seeing \citet{berg_etal_2022} for more details.

\subsection{Intermediate-Redshift Samples}
\label{sec:lowz}

Here we present the intermediate-redshift ($1\lesssim z\lesssim 4$) samples that we include in the analysis. 

We begin with the MOSFIRE Deep Evolution Field \citep[MOSDEF;][]{kriek_etal_2015} density sample presented in \citet{sanders_etal_2016}. The authors derive the electron density for galaxies at $2.0<z<2.6$ using both [\ion{O}{2}] and [\ion{S}{2}] at a median redshift of $z=2.24$. Of the total 69 galaxies in the MOSDEF density sample, 43 galaxies have densities derived using [\ion{O}{2}], and 26 have densities derived from [\ion{S}{2}]. We adopt the median [\ion{O}{2}] and [\ion{S}{2}] flux ratios reported by \citet{sanders_etal_2016} to rederive both of the median electron densities.

We also include the galaxy samples presented in \citet{kaasinen_etal_2017a} and \citet{strom_etal_2017}, who both derived $n_{\rm e}$ using the [\ion{O}{2}] doublet. \citet{kaasinen_etal_2017a} presented results from the COSMOS-[\ion{O}{2}] survey, which identified 21 galaxies at $1.4<z<1.7$ with robust detections of [\ion{O}{2}] and H$\alpha$ at a median redshift of $z=1.524$. \citet{strom_etal_2017} determined $n_{\rm e}$ for a subsample of 171 galaxies at $1.9<z< 2.7$, at a median redshift of $z=2.35$, as a part of the larger Keck Baryonic Structure Survey observations with MOSFIRE \citep[KBSS-MOSFIRE;][]{rudie_etal_2012, steidel_etal_2014, strom_etal_2017}. The authors compile $>3\sigma$ detections of the [\ion{O}{2}] doublet, and adopt the median [\ion{O}{2}] flux ratio to determine the median $n_{\rm e}$ for these galaxies.

The final intermediate-redshift sample that we include is presented in \citet{davies_etal_2021}, who measured $n_{\rm e}$ using [\ion{S}{2}] in four different sets of stacked galaxy spectra, binned by redshift: $z\leq0.1$ with 471 galaxies, $0.6<z<1.1$ with 39 galaxies, $1.1<z<1.9$ with 36 galaxies, and $1.9<z<2.6$ with 65 galaxies. The data for the $z\leq0.1$ stack is adopted from the SAMI survey \citep{croom_etal_2012, bryant_etal_2015}, while data for the remaining stacks are taken from the KMOS$^{\rm{3D}}$ survey \citep{wisinoski_etal_2015, wisinoski_etal_2019}. Both are integral field spectroscopic surveys, and we adopt the spatially integrated values for each galaxy.

\begin{figure}
\begin{center}
    \includegraphics[width=0.35\textwidth, trim=30 0 30 0,  clip=yes]{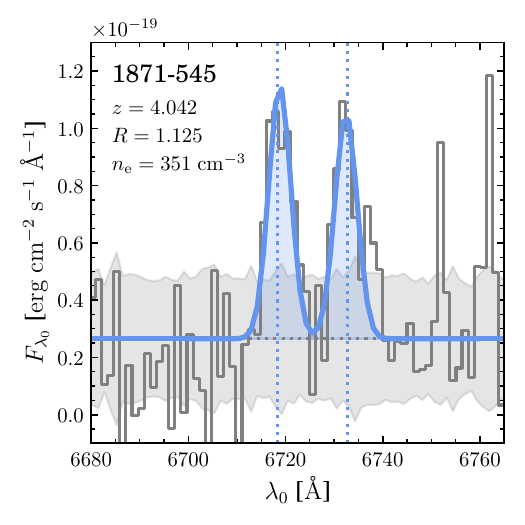}
        \caption{[\ion{S}{2}] $\lambda\lambda$6718, 6733\AA\AA~emission line profiles and Gaussian fits for the NIRSpec/MSA galaxy 1871-545. We provide the redshift of the galaxy, the measured [\ion{S}{2}] flux ratio, and the derived electron density. The rest-frame wavelengths of the [\ion{S}{2}]$\lambda\lambda$6718, 6733\AA\AA~emission lines are shown by vertical light blue dotted lines. The two doublet components for the Gaussian fits are shown  as shaded light blue, and the error on the data is shaded in grey. This galaxy is one of the highest redshift sources with densities measured from both [\ion{O}{2}] and [\ion{S}{2}], where $n_{\rm{e}}$[\ion{O}{2}$]=678^{+56}_{-40}~{\rm{cm}}^{-3}$ and $n_{\rm{e}}$[\ion{S}{2}$]=351^{+173}_{-53}~{\rm{cm}}^{-3}$.}
\label{fig:go_S2}
\end{center}
\end{figure}

\subsection{High-Redshift Samples}
\label{sec:highz}

Along with the local and intermediate-redshift galaxy samples, we include several samples containing galaxies at high-redshift ($z>4$) in our analysis, all consisting of observations with \textit{JWST}. 

We include electron density measurements presented by \citet{topping_etal_2025} for the Assembly of Ultradeep Rest-optical Observations Revealing Astrophysics \citep[AURORA; GO 1914; Co-PIs Shapley \& Sanders;][]{shapley_jwst_go1914, shapley_etal_2025a} survey. Observed with the NIRSpec medium-resolution grating, the entire survey has detections of the density-sensitive [\ion{S}{2}] doublet. The full sample is split into three smaller subsamples: 31 galaxies at $1.4<z<2.5$, 16 galaxies at $2.9<z<3.5$, and four galaxies at $4.4<z<5.8$.

We also include six additional $4 \lesssim z \lesssim 8$ galaxies with reported individual line fluxes from the [\ion{O}{2}]$\lambda\lambda$3727,3730\AA\AA~doublet to be included in the NIRSpec/MSA sample, as they were observed with the same filter-grating configurations as the GO 1871 program. These galaxies, presented in \citet{li_etal_2025}, were observed as a part of the JWST Early Release Science program \citep[ERS;][]{pontoppidan_etal_2022} Grism Lens-Amplified Survey from Space \citep[GLASS; DD-ERS 1324; PI Treu;][]{treu_etal_2022}. The program was designed to obtain deep high-resolution NIRSpec/MSA spectra of high-redshift galaxies lensed by the foreground galaxy cluster Abell 2744. They used the F170LP-G235H and F290LP-G395H filter-grating pairs, and exposure times were 4.9 hr for each filter-grating pair. We use the [\ion{O}{2}] doublet fluxes reported by \citet{li_etal_2025} to recalculate the densities to match our self-consistent approach. Combining these remaining GLASS targets and GO 1871 targets with resolved detections of [\ion{O}{2}] $\lambda\lambda$3727,3730\AA\AA~provides us with a sample of 12 galaxies, with a median redshift of $z=5.676$.

\section{Methods} 
\label{sec:methods}

Here we detail the methodology of our analysis, including measurements of emission lines and derived properties.

\subsection{Line Measurements} 
\label{sec:meas}

We measure all emission line fluxes and uncertainties for the NIRSpec/MSA and NIRCam/grism samples using the Python spectroscopic data analysis package {\texttt{SpecUtils}} \citep{astropy_2022} by fitting Gaussian profiles. Redshifts of each individual NIRSpec GO-1871 galaxy are determined using the strong [\ion{O}{3}]$\lambda$5008 Å line, and using the best-fit central wavelength we correct the observed wavelengths, flux densities and associated uncertainties for redshift to perform all line measurements in the rest-frame. When applying Gaussian fits to the emission lines, the width of each component is fixed to the velocity width of [\ion{O}{3}]$\lambda$5008 Å for the NIRSpec/MSA sample, and H$\alpha$ for the NIRCam/grism stacks. We do not constrain the amplitude or mean, but instead give {\texttt{SpecUtils}} a visually inspected peak flux and central wavelength as an initial estimate for the fitting process, which then returns the best-fit parameters. The doublets are fit simultaneously with a double component Gaussian. We report all properties derived using emission line measurements in Tables \ref{tab:nirspec_tab} and \ref{tab:nircam_tab}. All line fluxes used for computing the [\ion{O}{2}] and [\ion{S}{2}] diagnostic line ratios (\S\ref{sec:dens}) are not corrected for reddening, as differential extinction is negligible due to the close wavelength separation of each respective doublet emission line. All emission line fits for the NIRSpec/MSA and NIRCam/grism samples are presented in Figures \ref{fig:go_O2} and \ref{fig:congress_fresco_S2}, and the [\ion{S}{2}] doublet fit for 1871-545 in Figure \ref{fig:go_S2}.

\subsection{Electron Densities} 
\label{sec:dens}

\begin{figure}
\begin{center}
    \includegraphics[width=0.43\textwidth, trim=30 0 30 0,  clip=yes]{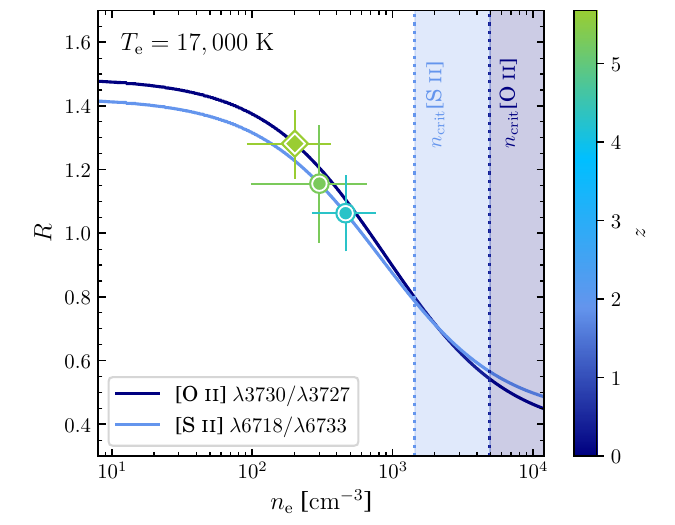}
        \caption{Diagnostic relation between measured flux ratio $R$ and resulting electron density $n_{\rm{e}}$ for [\ion{O}{2}] in dark blue and [\ion{S}{2}] in light blue, for our selection of atomic data. The sample medians are colored by median redshift. We derive this relation at our adopted electron temperature estimate, $T_{\rm{e}}=17000$ $\pm$ 500 K. For reference, we show the critical densities at $T_{\rm{e}}=17,000~\rm{K}$; $n_{\rm{crit}}$[\ion{O}{2}] shaded dark blue and $n_{\rm{crit}}$[\ion{S}{2}] shaded light blue, respectively.}
\label{fig:ne_R}
\end{center}
\end{figure}

To compute electron densities, we make use of the Python nebular analysis package {\texttt{pyneb}}\footnote{\url{https://github.com/Morisset/PyNeb_devel/tree/master}} \citep[][v.1.1.24]{luridiana_morriset_shaw_2013, luridiana_morisset_shaw_2015, morisset_etal_2020}. We first define a selection of atomic data: \citet{froese-fischer_tachiev_2004} and \citet{podobedova_etal_2009} for transition probabilities, as well as \citet{kisielius_etal_2009} and \citet{tayal_zatsarinny_2010} for collision strengths for O$^{+}$ and S$^{+}$, respectively. We also make use of the atomic data from \citet{storey_hummer_1995} for H$^0$ recombination lines.

For all [\ion{O}{2}] doublet measurements, we compute the diagnostic line flux ratio $R$[\ion{O}{2}], which we define as:
\begin{equation}
    R[\rm{O~{\textsc{ii}}}]=\frac{[\rm{O~{\textsc{ii}}}]\lambda3730Å}{[\rm{O~{\textsc{ii}}}]\lambda3727Å},
\label{eq:4}
\end{equation}
and the associated uncertainties, which are symmetric and derived from error propagation of the integrated flux uncertainties. The same treatment is given for [\ion{S}{2}] doublet measurements, with the diagnostic ratio R[\ion{S}{2}], which we define as:
\begin{equation}
    R[\rm{S~{\textsc{ii}}}]=\frac{[\rm{S~{\textsc{ii}}}]\lambda6718Å}{[\rm{S~{\textsc{ii}}}]\lambda6733Å}.
\label{eq:5}
\end{equation}
We then compute $n_{\rm{e}}$[\ion{O}{2}] and $n_{\rm{e}}$[\ion{S}{2}] using the \texttt{pyneb} task {\texttt{getTemDen}}, which returns either the temperature or density given the other for a defined diagnostic line ratio, as shown in Figure \ref{fig:ne_R}. For our density calculations, we fix the electron temperature at $T_{\rm{e}}=17,000$ $\pm$ 500 K, a typical electron temperature for galaxies at $z>4$ \citep[e.g.][]{topping_etal_2025}. This temperature is consistent with those derived for high-redshift galaxies \citep[e.g.][]{sanders_etal_2024, kumari_etal_2024, scholte_etal_2025}, and temperatures between $T_{\rm{e}}=10,000-20,000$ K are typically assumed for other density samples at high-redshift \citep[e.g.][]{isobe_etal_2023, abdurrouf_etal_2024, li_etal_2025, topping_etal_2025}. We find that temperature adjustments between $10,000-20,000$~K only have small ($<0.1$ dex) effects on $n_{\rm e}$ results, in agreement with \citet{topping_etal_2025}. We calculate the uncertainties on $n_{\rm{e}}$ by performing a Monte Carlo analysis, and varying the flux ratios and adopted temperature by their $1\sigma$ errors 1,500 times. These are reported as the 16th and 84th percentile values of the uncertainty distribution. Three galaxies in the NIRSpec/MSA sample have large $R$[\ion{O}{2}] values that surpass the high ratio limit ($R\geq1.5$, see Fig. \ref{fig:ne_R}). This corresponds to the low density limit, and returns unphysical or inconclusive results for $n_{\rm{e}}$[\ion{O}{2}]. For these galaxies, because they are all consistent with the low-density limit at $1\sigma$, we take the $1\sigma$ confidence value of the flux ratio and use it to recompute $n_{\rm{e}}$[\ion{O}{2}] as an upper limit. We also verify that none of our derived individual or median densities exceed the critical density $n_{\rm{crit}}$ of either [\ion{O}{2}] or [\ion{S}{2}] of $n_{\rm{crit}}$[\ion{O}{2}$]=4921~{\rm{cm}}^{-3}$ and $n_{\rm{crit}}$[\ion{S}{2}$]=1441~{\rm{cm}}^{-3}$ at our adopted $T_{\rm{e}}$.

Due to the nonlinear behavior of the diagnostic relationship between $R$ and $n_{\rm{e}}$, \citep[Fig. \ref{fig:ne_R}, and see e.g.][]{peng_etal_2025a}, it is imperative to carefully consider how this could affect the $n_{\rm e}$ derivation process. We can see in Figure \ref{fig:ne_R} that the diagnostic curve is insensitive to the $n_{\rm{e}}$ at low and high $R$, asymptotically approaching the theoretical maximum and minimum $R$, respectively. Because of this, a fairly symmetric distribution of $R$ leads to a very asymmetric distribution of $n_{\rm{e}}$. For these reasons, we follow the procedure outlined in past studies of typical electron densities at given redshifts \citep[e.g.][]{sanders_etal_2016, isobe_etal_2023}, and compute the median $R$ from the distribution of flux ratios derived per sample. The median $R$ uncertainties for the NIRSpec/MSA and NIRCam/grism samples are bootstrapped from the flux ratios and their associated uncertainties 1,500 times, and are reported as the 16th and 84th percentiles of the bootstrapped distribution. We then use this measurement to derive the median $n_{\rm{e}}$ for each sample. The median $n_{\rm{e}}$ uncertainties for all samples are bootstrapped from the median flux ratios and their associated uncertainties in the same manner as for the median $R$ uncertainties. For the NIRCam/grism stacked galaxy samples, we adopt the measured [\ion{S}{2}] doublet ratios as median ratios. 

\begin{figure}
\begin{center}
    \includegraphics[width=0.35\textwidth, trim=30 0 30 0,  clip=yes]{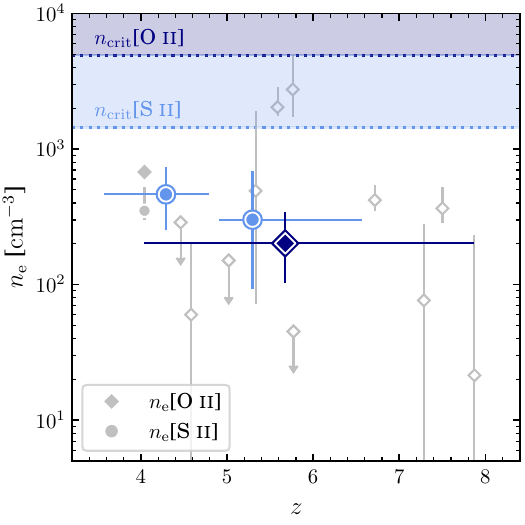}
        \caption{Individual and median electron densities and redshifts derived for the NIRSpec/MSA and NIRCam/grism samples. The individual galaxies in the NIRSpec/MSA sample are shown in light grey, and the median densities for each sample are shown in dark and light blue, for [\ion{O}{2}] and [\ion{S}{2}] densities, respectively. For reference, we show the critical densities at $T_{\rm{e}}=17,000~\rm{K}$; $n_{\rm{crit}}$[\ion{O}{2}] shaded dark blue and $n_{\rm{crit}}$[\ion{S}{2}] shaded light blue, respectively. We show the galaxy 1871-545 as two solid light grey points, corresponding to the [\ion{O}{2}] and [\ion{S}{2}] densities derived for that galaxy (see Figs. \ref{fig:go_O2} and \ref{fig:go_S2}).}
\label{fig:ne_z_smol}
\end{center}
\end{figure}

In addition, we simulate the nonlinear mapping of electron densities from each diagnostic ratio and add a correction to each individual Monte Carlo run when the density approaches the low-density limit. To do this, we generate 100,000 mock values of $R$[\ion{O}{2}] and $R$[\ion{S}{2}] between the range of validity (see Fig. \ref{fig:ne_R}) and compute the median $R$ and the resulting $n_{\rm{e}}$, for each tracer respectively. Then, we take the same mock values of $R$ and compute $n_{\rm{e}}$ for each realization, and compute the median $n_{\rm{e}}$, which we denote as the ``true" sample median. We choose the latter due to this median deviating the least from the actual peak of the distribution of both mock $n_{\rm{e}}$[\ion{O}{2}] and $n_{\rm{e}}$[\ion{S}{2}] values we compute. We compute the difference between these two medians and divide by the ``true" median, define this as the correction of the nonlinear mapping, and propagate the correction as an uncertainty into the bootstrapped median error by adding in quadrature. This effect is the most exaggerated for observed flux ratios near the low density limit, which is a common line ratio for many of our sources (for further discussion see \S\ref{sec:limits}). We present individual and median redshifts and densities for the NIRSpec/MSA sample, and median redshifts and densities for the NIRCam/grism samples in Figure \ref{fig:ne_z_smol}.

\subsection{Global Galaxy Properties} 
\label{sec:gal_props}

To investigate the dependency of electron density on other galaxy properties, we derive the color excess $E(B-V)$, the strong-line O$_{32}$ ratio, the total stellar mass ($M_\star$), the star-formation rate (SFR), the specific star-formation rate (sSFR), and the star-formation rate surface density ($\Sigma_{\rm{SFR}}$) for the NIRSpec/MSA and NIRCam/grism samples. We also include values derived for, and medians adopted from, the local, intermediate-, and high-redshift literature samples (Table \ref{tab:lit_tab}).

\subsubsection{Dust Extinction}
\label{sec:dust}

To correct all observed line fluxes for dust extinction in the NIRSpec GO 1871 sample, we use the line fluxes of the Balmer lines that meet the $\rm{S/N}\geq3$ criteria; H$\alpha$$\lambda$6564 Å, H$\beta$$\lambda$4862 Å, and H$\gamma$$\lambda$4341 Å. We compute all possible Balmer decrements for galaxies with two or more Balmer line detections, adopting the \citet{cardelli_etal_1989} extinction law. We then use {\texttt{pyneb}} to compute $E(B-V)$ and apply corrections to the line fluxes for each galaxy. We do this by using the {\texttt{getEmissivity}} task to compute the theoretical Balmer decrement, assuming Case B recombination \citep{dopita_sutherland_2003, groves_etal_2012}, adopting the same assumed electron temperature and derived electron densities as detailed in \S\ref{sec:dens}. We use tasks in {\texttt{RedCorr}} to determine $E(B-V)$, then derive and apply dust extinction corrections to measured line fluxes. Two of these galaxies, 1871-63 \citep{chisholm_etal_2024} and 1871-545 \citep{saldana-lopez_etal_2025}, have relatively to modestly low $E(B-V)$, while the remaining 4 galaxies all have negligible ($E(B-V)\sim0$) dust attenuation. For most of the remaining NIRSpec GLASS galaxies, we use the $E(B-V)$ values reported by \citet{nakajima_etal_2023}, who use the \citet{calzetti_etal_2000} extinction law. The source GLASS-160122 is dust-corrected using an alternate SED-applied version \citet{calzetti_etal_2000} law following the methods of \citet{theios_etal_2019}. For the NIRCam/grism CONGRESS and FRESCO galaxies, the grism only provides access to H$\alpha$ $\lambda$6564 Å, and have no derived $E(B-V)$ values. All derived parameters for the NIRCam/grism samples are not dust-corrected. Therefore, any interpretations involving these measurements must take this into consideration, as dust extinction curves of high-$z$ galaxies are shown to vary widely \citep[see e.g.][]{reddy_etal_2025, shapley_etal_2025b}.

The local literature samples are all dust-corrected using the same extinction law and process as the NIRSpec and NIRCam samples. The intermediate-redshift sample \citet{sanders_etal_2016} use the \citet{cardelli_etal_1989} extinction law, while \citet{strom_etal_2017} use the \citet{calzetti_etal_2000} law. AURORA galaxies are corrected for dust using the \citet{cardelli_etal_1989} extinction law. For more complete discussions regarding dust extinction for samples from the literature, we refer the reader to discussions in the intermediate- \citep{sanders_etal_2016, strom_etal_2017} and high-redshift \citep[AURORA;][]{topping_etal_2025} sample papers. Note that the \citet{cardelli_etal_1989} and \citet{calzetti_etal_2000} dust extinction laws are consistent in the rest-frame optical, so the difference is negligible.

\subsubsection{O$_{\textit{32}}$ Ratio}
\label{sec:o32}

To compute the O$_{32}$ ratio for the NIRSpec sample, which we define as:
\begin{equation}
    \rm{O_{32}}=\frac{[\rm{O~{\textsc{iii}}}]~\lambda5008}{[\rm{O~{\textsc{ii}}}]~\lambda3727+\lambda3730},
\label{eq:6}
\end{equation}
we first measure the [\ion{O}{3}]$\lambda$5008 Å line following the procedure outlined in \S \ref{sec:meas}, then compute O$_{32}$ using the [\ion{O}{2}] and [\ion{O}{3}] extinction-corrected intensities for each individual galaxy in the NIRSpec/MSA sample, and use these to compute the median O$_{32}$ of the NIRSpec/MSA sample. The median uncertainties are bootstrapped from the individual ratios with the same methods as above. For the NIRSpec/MSA GLASS targets, we adopt the Balmer dust-corrected O$_{32}$ measurements published by \citet{nakajima_etal_2023}. As for the NIRCam/grism sample stacks, we are not able to compute O$_{32}$, due to the grism not covering the [\ion{O}{2}] doublet or [\ion{O}{3}] rest-frame transitions.

We follow the same procedure for computing O$_{32}$ for the local samples (SDSS, LzLCS, CLASSY), and adopt the median O$_{32}$ from \citet{sanders_etal_2016} and AURORA for their density samples we include in this work. The remaining samples we include do not provide any published line fluxes or ratios that enable us to determine the O$_{32}$ ratio.

\subsubsection{Total Stellar Mass}
\label{sec:mstar}

The NIRSpec/MSA GO 1871 galaxies all have stellar masses derived using a customized version of the SED fitting code Bayesian Analysis of Galaxies for Physical Inference and Parameter EStimation \citep[\texttt{BAGPIPES}\footnote{\url{https://github.com/ACCarnall/bagpipes}};][]{carnall_etal_2018}, with \citet{bruzual_charlot_2003} stellar population synthesis templates, a \citet{kroupa_2001} initial mass function (IMF), and a \citet{calzetti_etal_2000} extinction law \citep[see][for more regarding SED fitting for the GO 1871 galaxies]{saldana-lopez_etal_2025}. For most of the NIRSpec/MSA GLASS galaxies, stellar masses were determined using \texttt{Prospector}\footnote{\url{https://github.com/bd-j/prospector}} \citep{johnson_etal_2019, johnson_etal_2021}, with the Flexible Stellar Population Synthesis \citep[\texttt{FSPS}\footnote{\url{https://github.com/cconroy20/fsps}};][]{conroy_gunn_white_2009, conroy_white_gunn_2010} templates, a \citet{chabrier_2003} IMF, and the \citet{calzetti_etal_2000} extinction law \citep[see][for regarding SED fitting for the GLASS galaxies]{nakajima_etal_2023}. The source GLASS-160122 has a stellar mass derived instead using \texttt{BAGPIPES}, with the 2016 \citet{bruzual_charlot_2003} stellar population synthesis templates, a \citet{chabrier_2003} IMF, and a \citet{calzetti_etal_2000} extinction law \citep[see][for more regarding SED fitting for GLASS-160122]{li_etal_2025}. We use these measurements to compute the median total stellar mass. Median $n_{\rm e}$ uncertainties are bootstrapped using the same methods as above.

For the NIRCam/grism sample datasets, total stellar masses are derived with \texttt{Prospector} using the \texttt{FSPS} templates a \citet{chabrier_2003} IMF, and a \citet{calzetti_etal_2000} extinction law. Stellar masses and their associated uncertainties are derived for each individual galaxy in CONGRESS and FRESCO, and we derive the median stellar mass and median uncertainties following the same methods that are used in other median uncertainty calculations. For a more complete discussion of the SED fitting process for CONGRESS and FRESCO sources, we refer the reader to the relevant sections in \citet{naidu_etal_2022} and \citet{tacchella_etal_2022}.

We also adopt median total stellar masses from the literature samples \citep{kauffmann_etal_2003, maraston_etal_2013, sanders_etal_2016, strom_etal_2017, flury_etal_2022, berg_etal_2022, topping_etal_2025}. These are determined using slightly different methodologies, but we ensure that all samples adopted the \citet{kroupa_2001} and \citet{chabrier_2003} IMFs. We derive the median total stellar masses for the local (SDSS DR7, SDSS DR16, LzLCS, CLASSY) and AURORA literature samples following the procedure for the NIRSpec/MSA galaxies. 

% For the SDSS sample, stellar masses are derived following the SED fitting methods described in \citet{kauffmann_etal_2003} and \citet{salim_etal_2007}, which use the \citet{bruzual_charlot_2003} stellar population synthesis templates and a \citet{chabrier_2003} IMF. The stellar masses for the LzLCS sample are derived using \texttt{Prospector} with a \citet{kroupa_2001} IMF and the \citet{calzetti_etal_2000} extinction law. Stellar masses are determined for the CLASSY sample using the BayEsian Analysis of GaLaxy sEds \citep[\texttt{BEAGLE}\footnote{\url{https://www.iap.fr/beagle/index.html}};][v.0.24.0]{chevallard_charlot_2016, gutkin_charlot_bruzual_2016} SED fitting code, with \citet{bruzual_charlot_2003} stellar population synthesis templates. For the density sample of \citet{sanders_etal_2016}, they derive their median total stellar mass using the SED fitting code \texttt{FAST}\footnote{\url{https://github.com/jamesaird/FAST}} \citep{kriek_etal_2009}, with the \texttt{FSPS} templates, a \citet{chabrier_2003} IMF, and assumes the \citet{calzetti_etal_2000} extinction law. \citet{strom_etal_2017} determine their median total stellar mass using SED fitting techniques with \citet{bruzual_charlot_2003} stellar population synthesis templates, a \citet{chabrier_2003} IMF, and uses the \citet{calzetti_etal_2000} extinction law. Lastly, the AURORA density sample derives total stellar masses using SED fitting with \texttt{FAST}, assuming the \texttt{FSPS} templates and a \citet{chabrier_2003} IMF. 

\subsubsection{Star-Formation Rate}
\label{sec:sfr}

To compute the star-formation rate (SFR) for the NIRSpec/MSA and NIRCam/grism galaxy samples, we use the \citet{kroupa_2001} IMF-adjusted values for the \citet{kennicutt_evans_2012} SFR law, defined as:
\begin{equation}
    \log({\rm{SFR_{H\alpha}}})=\log({L_{\rm{H\alpha}}})-41.27
\label{eq:7}
\end{equation}
and use \texttt{pyneb} and the redshift to determine the dust-corrected H$\alpha$ luminosity. For the NIRSpec/MSA GLASS sources, \citet{nakajima_etal_2023} use the H$\beta$ luminosity and the \citet{kennicutt_1998} SFR law, with conversion factors from \citet{madau_dickinson_2014} applied to the \citet{chabrier_2003} IMF. For the source GLASS-160122, \citet{li_etal_2025} use the \citet{kennicutt_1998} SFR law and a \citet{chabrier_2003} IMF. We use these individual NIRSpec/MSA values to compute the median SFR. The median uncertainties are bootstrapped in the same manner that we did for the rest of the samples. Note that the NIRCam/grism sample values for the H$\alpha$ luminosity are not dust corrected, as stated above in \S\ref{sec:dust}.

The local literature samples (SDSS DR7, SDSS DR16, LzLCS, CLASSY) all have median SFRs derived using the same process as for the NIRSpec GO 1871 galaxies. The \citet{sanders_etal_2016} and \citet{strom_etal_2017} samples use the \citet{kennicutt_1998} SFR law to derive their median SFRs. As for AURORA, the median SFRs are also determined following the procedure outlined for the NIRSpec/MSA GO 1871 galaxies.

\subsubsection{Specific Star-Formation Rate}
\label{sec:ssfr}

The specific star-formation rate (sSFR) traces the rate of star-formation per stellar mass, and is defined as:
\begin{equation}
    \log({\rm{sSFR})=\log({\rm{SFR}}})-\log({M_\star}).
\label{eq:8}
\end{equation}
We use the individual NIRSpec/MSA sample values to compute the median sSFR. The median uncertainties are bootstrapped from the individual ratios and their associated uncertainties using the same methods as above. We also take the median total stellar masses and H$\alpha$-derived SFRs from the local, intermediate- and high-redshift literature samples \citep[SDSS DR7, SDSS DR16, LzLCS, CLASSY,][AURORA]{sanders_etal_2016, strom_etal_2017} to derive their median sSFRs using the same derivation schema.

\subsubsection{Star-Formation Rate Surface Density}
\label{sec:sigma_sfr}

We define the star-formation rate surface density ($\Sigma_{\rm{SFR}}$) as:
\begin{equation}
    \Sigma_{\rm{SFR}}=\frac{\rm{SFR}}{2\pi r_e^2},
\label{eq:9}
\end{equation}
where $r_e$ is the effective radius of the source. \citet{saldana-lopez_etal_2025} determines $r_e$ for the NIRSpec/MSA GO 1871 galaxies by using \texttt{pysersic}\footnote{\url{https://github.com/pysersic/pysersic}} \citep{pasha_miller_2023} to fit 2D Sérsic profiles in the F444W photometry, and \citet{naidu_etal_2022} uses \texttt{GALFIT}\footnote{\url{https://users.obs.carnegiescience.edu/peng/work/galfit/galfit.html}} \citep{peng_etal_2002, peng_etal_2010} to fit 2D Sérsic profiles in the F444W CONGRESS and FRESCO images. The median uncertainties are bootstrapped with the same methods as above. As there are no published size measurements for any of our NIRSpec/MSA GLASS targets, we do not derive $\Sigma_{\rm{SFR}}$ for these galaxies. Therefore, the resulting median only includes half of the NIRSpec/MSA sample.

The local literature samples (LzLCS, CLASSY) with reported measurements of $r_e$ are derived similarly to the NIRSpec/MSA and NIRCam/grism samples, and for the high-redshift sample AURORA we adopt the values that are reported in \citet{topping_etal_2025}.

\section{Results}
\label{sec:results}

In this section, we report the main results of our study, including evidence of an evolution in electron density with redshift, with the ionization parameter, and no evidence of any correlations with other global galaxy properties.

\subsection{Median Electron Densities at $z\sim4-8$}
\label{sec:den_results}

We report the results from the {\textit{JWST}} NIRSpec/MSA and NIRCam/grism electron density samples here. We find a median [\ion{O}{2}] ratio of $R$[\ion{O}{2}$]=1.282^{+0.096}_{-0.106}$ for the NIRSpec/MSA sample at $\langle z\rangle = 5.6$, corresponding to a median electron density of $n_{\rm{e}}$[\ion{O}{2}$]=202^{+145}_{-102}~{\rm{cm}}^{-3}$. All measured and derived parameters for the individual NIRSpec/MSA galaxies, as well as the medians, are provided in Table \ref{tab:nirspec_tab}, though we note that a significant portion of sources have densities consistent with the low-density limit (as noted in Table \ref{tab:nirspec_tab} and Figure \ref{fig:ne_z_smol}). We find that median [\ion{S}{2}] ratios for the NIRCam/grism CONGRESS and FRESCO samples  at $\langle z\rangle = 4.2$ and $5.3$, respectively, are $R$[\ion{S}{2}$]=1.063\pm0.120$ and $R$[\ion{S}{2}$]=1.156\pm0.186$, which correspond to median electron densities of $n_{\rm{e}}$[\ion{S}{2}$]=463^{+281}_{-203}~{\rm{cm}}^{-3}$ and $n_{\rm{e}}$[\ion{S}{2}$]=301^{+388}_{-206}~{\rm{cm}}^{-3}$, respectively. All measured and derived medians for the NIRCam/grism datasets are provided in Table \ref{tab:nircam_tab}. These results are all presented in Figure \ref{fig:ne_z_smol}; the individual NIRSpec/MSA galaxies are shown in gray points with the medians of the CONGRESS, FRESCO and NIRSpec/MSA samples shown as light blue and dark blue points. The density medians are also presented in Figure \ref{fig:ne_z}.

We compare the results of our datasets, and find that all three median electron densities are consistent within their 1$\sigma$ uncertainties, even when using both [\ion{O}{2}] and [\ion{S}{2}] to determine densities, aligning with what is seen at lower redshifts \citet{sanders_etal_2016}. Conversely, one NIRSpec galaxy 1871-545 has electron densities derived from both [\ion{O}{2}] and [\ion{S}{2}], with $R$[\ion{O}{2}$]=1.005\pm0.076$ and $R$[\ion{S}{2}$]=1.125\pm0.311$, yielding $n_{\rm{e}}$[\ion{O}{2}$]=678^{+56}_{-40}~{\rm{cm}}^{-3}$ and $n_{\rm{e}}$[\ion{S}{2}$]=351^{+173}_{-53}~{\rm{cm}}^{-3}$. At $z=4.042$, this is one of the highest redshift sources with multiple low-ionization density diagnostics. We determine that these densities agree within 2$\sigma$, suggesting that they are largely able to trace the same regions of ionized gas in this galaxy \citep[e.g.][]{mendez-delgado_etal_2023b, peng_etal_2025a}.

We also compare our results with other galaxy samples published in the literature at similar redshifts. \citet{li_etal_2025} and \citet{topping_etal_2025} find median densities that are $\sim100~\rm{cm}^{-3}$ larger than our results, though we note that \citet{li_etal_2025} estimate their median without distinguishing between or accounting for the two different diagnostic ratios, and both studies lack the number of objects constraining our measurements. The median density determined by \citet{harikane_etal_2025b} using optical $R$[\ion{O}{2}] measurements is also larger than our results, yielding a value of $n_{\rm{e}}$[\ion{O}{2}$]\sim700~{\rm{cm}}^{-3}$, yet their sample median only relies on density measurements of four galaxies. \citet{reddy_etal_2023b} instead find a median [\ion{S}{2}] density $\sim1~\rm{dex}$ lower than our results, and we note that this median was based on [\ion{S}{2}] line measurements of a composite spectrum (see \citet{reddy_etal_2023b} for more details). Therefore, for comparison we include the \citet{reddy_etal_2023b} median $n_{\rm e}$[\ion{S}{2}] that we rederive using their reported median $R$[\ion{S}{2}] (following the analysis in \S\ref{sec:dens}, $n_{\rm e}$[\ion{S}{2}$]=62^{+115}_{-62}~\rm cm^{-3}$) in Figure \ref{fig:ne_z}. All four medians have uncertainties that are consistent with our JWST NIRSpec/MSA and NIRCam/grism median uncertainties.

We then compare our findings to other studies examining typical electron densities in $z<4$ galaxies, and find that our results are consistent with median electron densities found at $z\sim2$ \citep[e.g.][]{kaasinen_etal_2017a, strom_etal_2017, topping_etal_2025, raptis_etal_2025, rogers_etal_2025}, and are found to be $\sim1$ dex larger than those found at $z\sim0$. Given that we also rederive median electron densities for galaxy density samples from the literature, we find that our rederived median densities are systematically $\sim50-100~{\rm{cm}}^{-3}$ larger than what is reported in the literature, due to our chosen suite of atomic data (\S\ref{sec:dens}). The increase is observed in both $n_{\rm{e}}$[\ion{O}{2}] and $n_{\rm{e}}$[\ion{S}{2}].

\subsection{Evolution of Electron Density with Redshift}
\label{sec:z_den_evo}

\begin{figure*}
\begin{center}
    \includegraphics[width=0.8\textwidth, trim=30 0 30 0,  clip=yes]{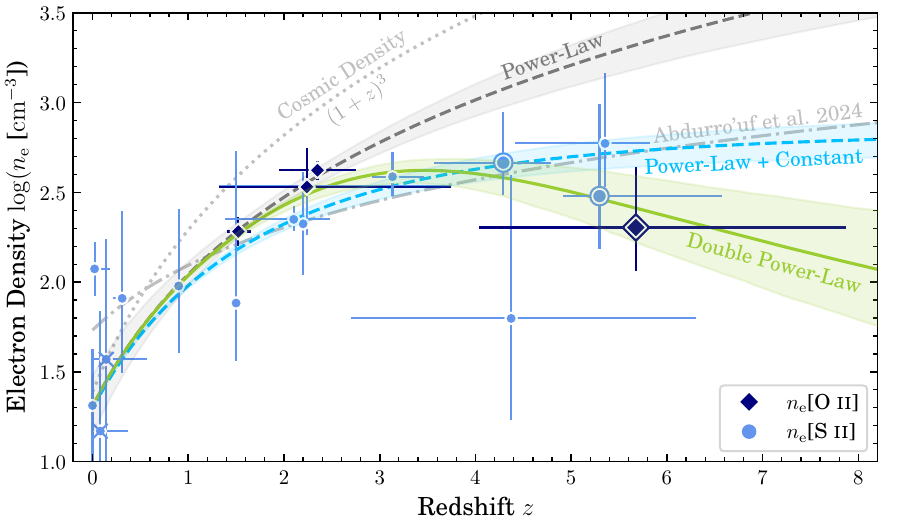}
        \caption{Evolution of low-ionization median electron densities derived with the density-sensitive [\ion{O}{2}] and [\ion{S}{2}] doublets with median redshift. The outlined data points indicate the NIRSpec/MSA and NIRCam/grism medians presented in this work, while all others are medians adopted for galaxy samples from the literature. The median densities for each sample are shown in dark and light blue, for [\ion{O}{2}] and [\ion{S}{2}] densities, respectively. The best-fit power law is shown as a dotted light-grey trend with shaded grey error, the best-fit power law with a flat evolution is shown as a dashed grey line, and the best-fit double-power law is shown in green with shaded green error (see \S\ref{sec:z_den_evo} for more details on the fitting procedure). We also include the best-fit power law from \citet{abdurrouf_etal_2024}. The SDSS medians are demarcated with an ``x''. Our combined medians are best fit by a double-power law evolution, contrary to other similar works at high-$z$.}
\label{fig:ne_z}
\end{center}
\end{figure*}

We then investigate trends of low-ionization electron densities determined with [\ion{O}{2}] and [\ion{S}{2}]. Figure \ref{fig:ne_z_smol} showcases the individual and median values of the NIRSpec/MSA sample, and the medians from the NIRCam/grism samples. As mentioned in \S\ref{sec:den_results}, though our results are within uncertainties of other findings at similar redshifts, we find that our median densities marginally decrease as median redshift increases beyond $z \sim 4$. Figure \ref{fig:ne_z} presents the NIRSpec/MSA and NIRCam/grism medians, as well as the literature medians from \S\ref{sec:lit} across redshift.

\begin{deluxetable*}{cccccccccccc}
\tablecaption{Redshift Evolution Fit Parameters and Statistics}
\tabletypesize{\scriptsize}
\tablewidth{0pt}
\tablehead{
\colhead{Fit Law} & \colhead{$a$} & \colhead{$b$} & \colhead{$c$} & \colhead{$d$} & \colhead{$\chi^2_{\nu}$} & \colhead{$\chi^2_{\nu}-\chi^2_{{\nu},\rm{DPL}}$} & \colhead{BIC} & \colhead{$\rm{BIC}-BIC_{\rm{DPL}}$}
}
\startdata
\citet{abdurrouf_etal_2024} & $1.73$ & $1.2$ & -- & -- & 3.53$\pm$1.49 & 2.77$\pm$1.52 & 65.9$\pm$25.3 & 47.1$\pm$25.9 \\
Power-Law & $1.31\pm0.16$ & $2.44\pm0.34$ & -- & -- & 13.4$\pm$7.42 & 12.6$\pm$7.43 & 233$\pm$126 & 214$\pm$126 \\
Power-Law+Constant & $1.31\pm0.16$ & $2.44\pm0.34$ & $2.44\pm0.38$ & $3.72\pm0.46$ & 1.24$\pm$0.57 & 0.48$\pm$0.66 & 25.3$\pm$10.2 & 6.42$\pm$11.7 \\
Double Power-Law & $1.31\pm0.16$ & $2.44\pm0.34$ & $5.22\pm1.08$ & $4.58\pm0.29$ & 0.76$\pm$0.34 & -- & 18.9$\pm$5.73 & -- \\[0.25pc]
\enddata
\tablecomments{Here $\chi^2_{\nu,\rm{DPL}}$ and $\rm BIC_{DPL}$ represent the $\chi^2_{\nu}$ and BIC for the double-power law (DPL) fit.}
\label{tab:stats_tab}
\end{deluxetable*}

We first check that there is a significant correlation between median $z$ and $n_{\rm{e}}$ by computing the Kendall's $\tau$ coefficient \citep{akritas_siebert_1996}, using \texttt{kendall}\footnote{\url{https://github.com/sflury/kendall}} \citep{flury_etal_2022, flury_kendalltau}. When accounting for outliers and uncertainties, we find that $\tau=0.68\pm0.10$ and $p=8.2\times10^{-5}$, which implies a strong likelihood of a correlation. Then, we fit a number of trends to the medians while accounting for median uncertainties using the \texttt{scipy}\footnote{\url{https://scipy.org}} \citep{scipy_2020} routine \texttt{curve\_fit}. We begin by fitting a single power-law function to the local and intermediate-redshift samples to establish a baseline of the trend up to $z\sim4$. We do this for three reasons: 1) The evolution of low-ionization $n_{\rm{e}}$ with $z$ at $0<z<4$ has been shown to match the behavior of a single power-law \citep[see e.g.][]{davies_etal_2021}, 2) the local and intermediate-redshift galaxy samples provide a basis for the trend from $0<z<4$, and 3) we fix the fit parameters from this low-$z$ fit as a starting point for performing the other fits. We find a best-fit power-law for the $z<4$ data:
\begin{equation}
    \log(n_{\rm{e}})=a+b\log(1+z),
\label{eq:10}
\end{equation}
where $a=1.39\pm0.03$ and $b=2.26\pm0.13$. Physically, $a$ is the median $n_{\rm{e}}$ at $z=0$, and $b$ is the rate at which median $n_{\rm{e}}$ grows with median $z$. This fit predicts that typical $n_{\rm{e}}$ at $z>4$ should be $\sim0.5$ dex larger than those at $z\sim2.5$. However, the observed NIRSpec and NIRCam median $n_{\rm{e}}$ suggest a break from a single power-law evolution. The single power-law closely matches the evolution of the samples up to $z\sim2.5$. At higher $z$ we find that the single power-law starts to overpredict the electron density evolution, and by $z\sim6$ our NIRCam/grism CONGRESS, FRESCO and NIRSpec/MSA median $n_{\rm{e}}$ values are increasingly discrepant from the single power-law ($1.2\sigma$, $1.6\sigma$, and $2.9\sigma$, respectively). Thus, there appears to be a change in the density evolution that occurs near $z\sim3$, which would be the first such evolutionary behavior observed in the low-ionization ISM at high-$z$. To examine this possibility, we fit a double power-law to our data and fix $a$ and $b$ to the fit coefficients found above (Eq. \ref{eq:10}). In this approach, one power-law describes the low-$z$ density evolution (Eq. \ref{eq:10}), while the second power-law describes the density evolution at high-$z$:
\begin{equation}
    \log(n_{\rm{e}})=a+b\log(1+z)-\log\left[1+\left(\frac{(1+z)}{d}\right)^c~\right].
\label{eq:11}
\end{equation}
We fit for a flat evolution of beyond $z\sim4$, letting $c=b$, and find a best-fit law where $d=4.47\pm0.54$. Physically, $c$ is the rate at which the median $n_{\rm{e}}$ falls with $z$ and $d$ is the turnover $1+z$, or the redshift beyond which the second term dominates the evolution. Then, we consider the possibility of a falling evolution of $n_{\rm{e}}$ with increasing $z$ beyond $z\sim4$ by letting both $c$ and $d$ vary. We find a best-fit double power-law where $c=5.21\pm1.15$ and $d=4.69\pm0.31$, with a peak in electron density of $n_{\rm{e}}\sim400$ cm$^{-3}$ at $z\sim3.5$. We also include the single power-law fit reported by \citet{abdurrouf_etal_2024} to compare against our results. All trends and median $n_{\rm{e}}$ across $z$ are shown in Figure \ref{fig:ne_z}. When comparing all of these fits, we find that there is a statistical preference for an alternative redshift evolution that does not follow a single power-law evolution. However, when looking at the flat and decreasing double power-law fits, we find no significant difference between the preference for these two models. These findings are supported by a reduced $\chi^2$ ($\chi^2_{\nu}$) analysis and the Bayesian Inference Criterion (BIC) values. The uncertainties on both the $\chi^2_{\nu}$ values and BICs are derived by bootstrapping the errors on the median $n_{\rm{e}}$ measurements and the best-fit parameters. The statistics and their uncertainties for each of the included fits are reported in Table \ref{tab:stats_tab}. We discuss implications of this alternate evolution in \S\ref{sec:discuss}.

\begin{figure}
\begin{center}
    \includegraphics[width=0.4\textwidth, trim=30 0 30 0,  clip=yes]{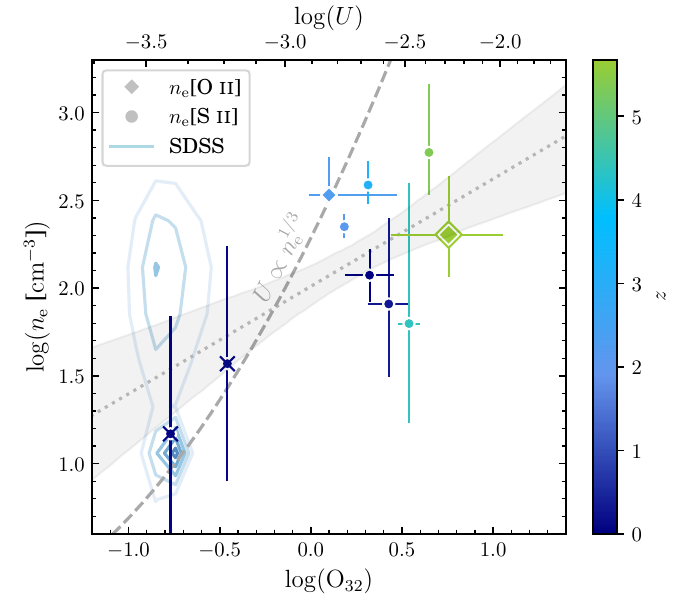}
        \caption{Evolution of median electron densities with median O$_{32}$ ratios. We include an upper x-axis for the ionization parameter $\log(U)$, parameterized by the relations of \citet{berg_etal_2019} and \citet{berg_etal_2024}. The sample medians are colored by median redshift. We show the best-fit trend between $n_{\rm{e}}$ and O$_{32}$ (Eq. \ref{eq:12}) as a dotted light grey line, with shaded grey error. For reference, we also include the theoretical relation where $U\propto n_{\rm{e}}^{1/3}$, shown as a dashed grey line. The SDSS medians are demarcated with an ``x'', and the full SDSS samples are shown as blue contours. Our combined datasets do not match the theoretical relation between $U$ and $n_{\rm e}$ for a radiation-bounded system, having a shallower evolution than expected.}
\label{fig:ne_O32_logU}
\end{center}
\end{figure}

\begin{figure*}
\begin{center}
    \includegraphics[width=\textwidth, trim=30 0 30 0,  clip=yes]{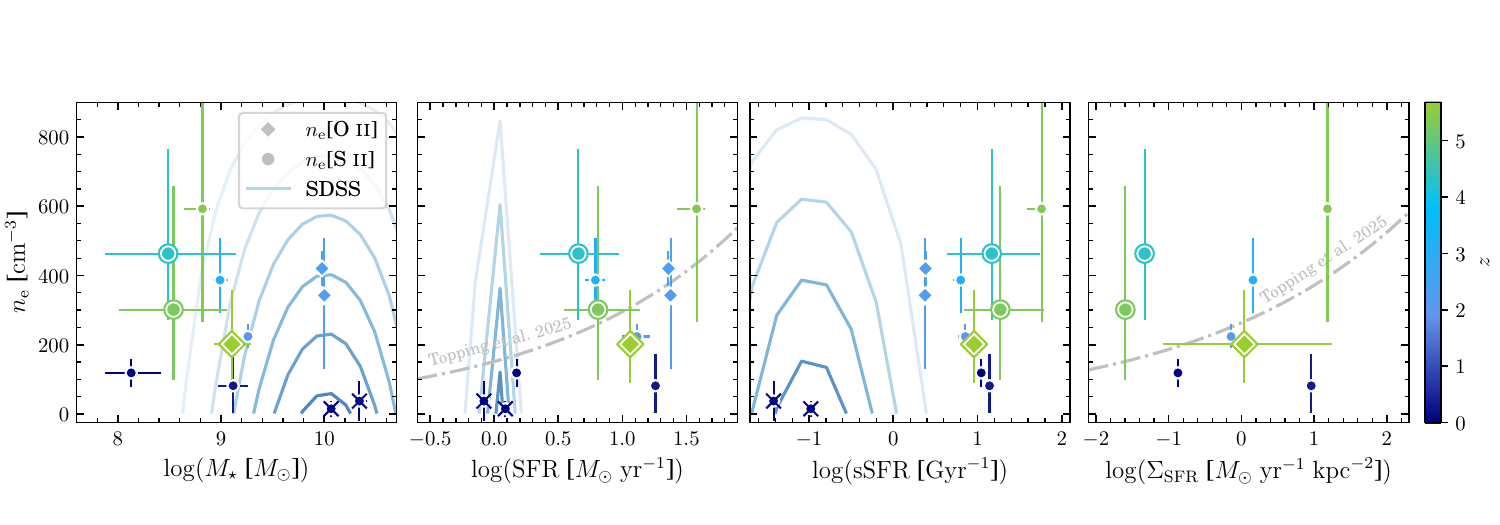}
        \caption{Median electron densities as a function of median galaxy properties. \textit{From left to right;} median total stellar mass, median SFR, median sSFR, and median $\Sigma_{\rm{SFR}}$. The data points are colored by median redshift. While we do not find statistically significant relations between the data points, we include the best-fit relations for SFR and $\Sigma_{\rm{SFR}}$ as a function of electron density from \citet{topping_etal_2025}, shown as a dash-dotted grey line. The SDSS medians are demarcated with an ``x'', and the full SDSS samples are shown as blue contours. We cannot confirm any strong trends between $n_{\rm{e}}$ and any of these galaxy properties with our combined datasets in this work.}
\label{fig:ne_props}
\end{center}
\end{figure*}

\subsection{Evolution of Electron Density with Ionization Parameter}
\label{sec:ion_den_evo}

With our collected measurements of the median O$_{32}$ ratios across redshift, we also investigate the possibility of an evolutionary trend with the median electron density. The O$_{32}$ ratio can be used to derive the ionization parameter $U$, such that we can trace how the typical ionization states and structures of galaxies change across cosmic time. To examine this directly, we plot the median O$_{32}$ ratios of the NIRSpec/MSA, NIRCam/grism, and literature density samples in Figure \ref{fig:ne_O32_logU}. The O$_{32}$ median values are also converted to $\log(U)$ following the parameterizations presented in \citet{berg_etal_2019} and \citet{berg_etal_2024}, determined from the photoionization modeling software {\texttt{Cloudy}} \citep{ferland_etal_2013}. We assume solar metallicity \citep[$Z=Z_\odot$][]{berg_etal_2024} for the local SDSS DR7 and DR16 density samples, and we assume 20\% solar \citep[$Z=0.2Z_\odot$][]{berg_etal_2019} for the local analog, intermediate-, and high-$z$ density samples. Given that we are unable to derive the median metallicity of the NIRSpec/MSA and NIRCam/grism samples, we also assume 20\%~Z$_\odot$ is a typical metallicity for these samples at their respective median redshifts \citep{curti_etal_2023, sanders_etal_2024}. Note that at a fixed $\log(\rm O_{32})>0$ the choice of metallicity between solar and 70\% solar can vary inferred $\log(U)$ by $\sim0.1$ dex, and that at fixed $\log(\rm O_{32})<0$ the choice of metallicity between 40\% solar and 2\% solar can vary inferred $\log(U)$ by $\sim0.5$ dex \citep[see][for more details]{berg_etal_2019, berg_etal_2024}. With that in mind, we find a positive correlation with $n_{\rm{e}}$ and O$_{32}$, which implies a similar correlation with $U$. To quantify this relation, we fit a trend to the median $\log($O$_{32})$ and $\log(n_{\rm{e}})$ values:
\begin{equation}
    \log(n_{\rm{e}})=a+b\log(\rm{O}_{32}),
\label{eq:12}
\end{equation}
and find the best-fit trend where $a=2.01\pm0.12$ and $b=0.61\pm0.26$. For comparison, we also include the theoretical relation (Eq. \ref{eq:3}) between $U$ and $n_{\rm{e}}$ for a radiation-bounded \ion{H}{2} region, assuming that the rate of ionizing photons $Q$ and the volume filling fraction $\epsilon$ are both constant (Fig. \ref{fig:ne_O32_logU}). We do acknowledge that this may be due to the fact that our metallicity assumption likely does not apply to all the data (especially to SDSS galaxies), and therefore may be underestimating $U$ by $\sim0.5$ dex. While there is a trend between $z$ and O$_{32}$, the stronger trend appears to be with O$_{32}$ as low-$z$ points (such as CLASSY and LzLCS+) reside at similar $n_{\rm{e}}$ as higher $z$ galaxies with similar O$_{32}$. We also include our rederived \citet{reddy_etal_2023b} median $n_{\rm e}$[\ion{S}{2}] and O$_{32}$ median (following the analysis in \S\ref{sec:o32}) in Figure \ref{fig:ne_O32_logU} for comparison, and use the \citet{berg_etal_2019} schema assuming 20\% solar to convert to $\log(U)$. We find that the \citet{reddy_etal_2023b} medians match our trend within uncertainty. We discuss these findings further and other implications in \S\ref{sec:discuss}.

\citet{reddy_etal_2023b} also find that O$_{32}$ (and therefore $U$) and $n_{\rm{e}}$ scale with each other, following the expectations for a radiation-bounded nebula. They also examine whether or not the ionizing photon production rate $Q$ is a significant driver of $U$, but did not find a significant difference between $Q$ for low-O$_{32}$ and high-O$_{32}$ galaxies, which they suggest may be due to the limited dynamic range in SFR probed by their sample. However, \citet{reddy_etal_2023a} suggest that an increasing $n_{\rm{e}}$ and $Q$ (as probed by SFR) may be responsible for $U$ increasing with $z$. We test this by including SFR in a multivariate fit, but do not find a statistically significant correlation. However, we cannot definitively rule out the possibility that $Q$ is a significant driver of $U$ at $z>4$.

\subsection{Evolution of Electron Density with Global Galaxy Properties}
\label{sec:galprops_den_evo}

\begin{figure*}
\begin{center}
    \includegraphics[width=0.7\textwidth, trim=30 0 30 0,  clip=yes]{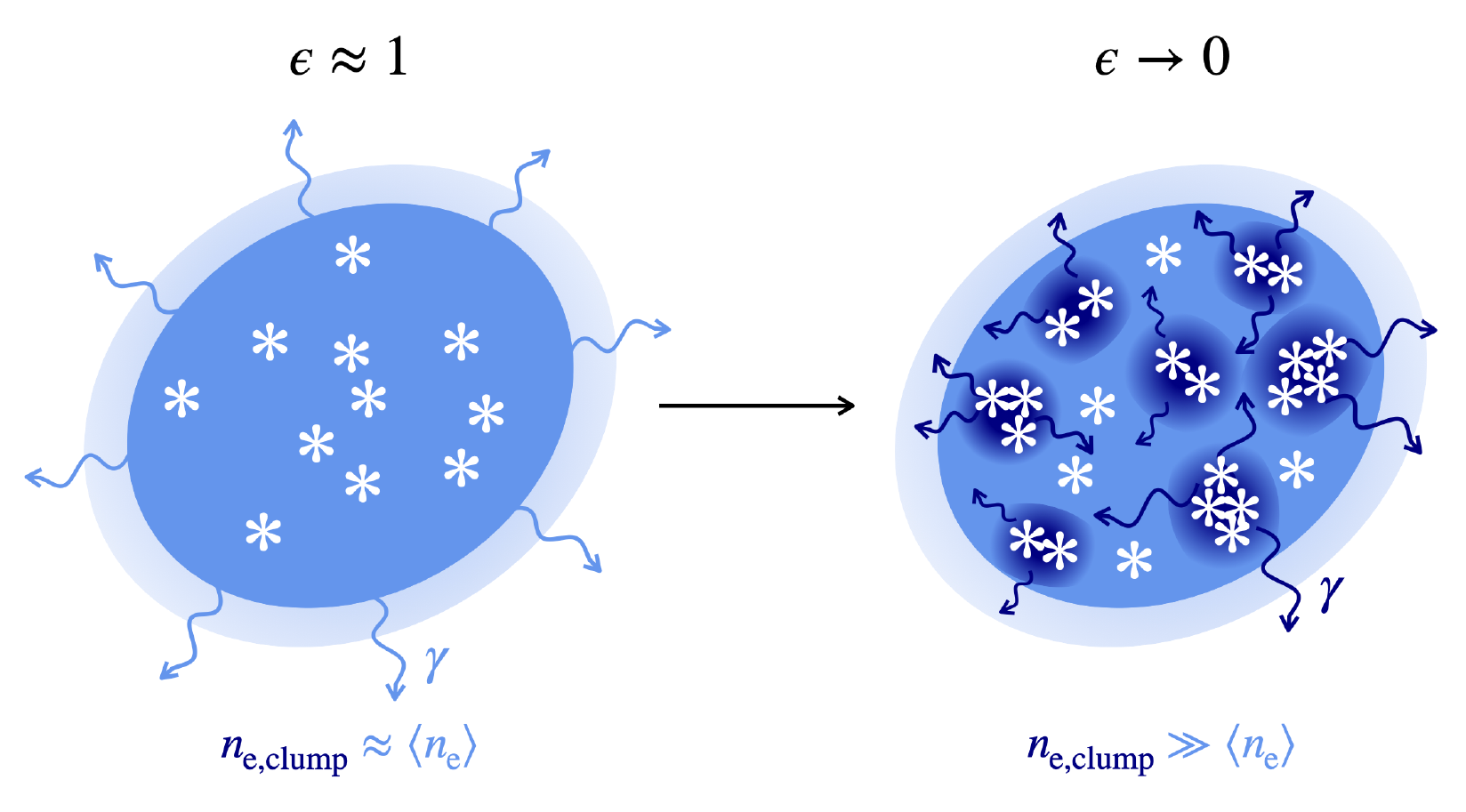}
        \caption{Schematic illustration of our proposed inhomogeneous gas density model. \textit{Left:} When $\epsilon\approx1$, the clump density $n_{\rm{e, clump}}$ and the mean volume-averaged density $\langle n_{\rm{e}} \rangle$ are roughly equivalent. This would result in both regions having similar emission line signatures from density-sensitive tracers, such as [\ion{O}{2}] and [\ion{S}{2}]. Assuming $\langle n_{\rm{e}} \rangle$ is within the valid sensitivity range of both tracers, both $n_{\rm{e}}$[\ion{O}{2}] and $n_{\rm{e}}$[\ion{S}{2}] directly probe $\langle n_{\rm{e}} \rangle$. \textit{Right:} When $\epsilon\rightarrow0$, $n_{\rm{e, clump}}$ becomes much larger than $\langle n_{\rm{e}} \rangle$. The emission is more concentrated in clumps, meaning that [\ion{O}{2}] and [\ion{S}{2}] emission signatures are luminosity weighted towards these clumps. Assuming $n_{\rm{e, clump}}$ is within the valid sensitivity range of both tracers, both $n_{\rm{e}}$[\ion{O}{2}] and $n_{\rm{e}}$[\ion{S}{2}] are more sensitive to $n_{\rm{e, clump}}$, and cannot reliably trace $\langle n_{\rm{e}} \rangle$. This implies that densities measured from emission lines will always bias towards line-emitting structures in an ionized region.}
\label{fig:neb_model}
\end{center}
\end{figure*}

We also search for trends with median electron density and other median galaxy properties; $M_\star$, SFR, sSFR, and $\Sigma_{\rm{SFR}}$. To examine possible trends, we plot these results in Figure \ref{fig:ne_props}, where we see evidence of tentative trends with median $n_{\rm{e}}$ and both median $M_\star$ and sSFR. To test the likelihood of a correlation, we compute the $p$-value for each panel using the \texttt{kendall}'s $\tau$ test, and find little indication of any strong trends (all $p$-values are $>0.1$) with any of these galaxy properties and $n_{\rm e}$. This is likely due to the small number of medians that have reliable and consistent measurements. Other authors have also found little evidence of strong trends between $n_{\rm{e}}$ and these global galaxy properties \citep[e.g.][]{li_etal_2025, topping_etal_2025}, and suggest similar observational constraints cannot be ruled out before drawing any significant conclusions. One exception to this is evidence in the literature of a significant correlation between $n_{\rm{e}}$ and $\Sigma_{\rm{SFR}}$ \citep[e.g.][]{shimakawa_etal_2015, davies_etal_2021, reddy_etal_2023a, reddy_etal_2023b}. Are results are consistent with these previous relations, but we do not find a statistical trend between $n_{\rm e}$ and $\Sigma_{\rm SFR}$. However, this is still based on only a few galaxy samples at limited redshift ranges, and the scatter in the relation is large. Each of these trends with $n_{\rm{e}}$ requires more homogeneous observations of diverse populations of galaxies across redshift to confirm.

\section{Discussion}
\label{sec:discuss}

In \S\ref{sec:den_results} and \S\ref{sec:z_den_evo}, we find that the median $n_{\rm{e}}$ derived using the optical [\ion{O}{2}] and [\ion{S}{2}] diagnostics appear to deviate from a single power-law evolution with redshift beyond $z\sim4$. We also find in \S\ref{sec:ion_den_evo} that the median $n_{\rm{e}}$ appears to increase with the median O$_{32}$ (and $U$), although at a slower rate than expected from theory (Eq. \ref{eq:3}). We then attempt to examine what may be driving these trends by looking at how median $n_{\rm{e}}$ scales with median $M_\star$, SFR, sSFR, and $\Sigma_{\rm{SFR}}$ in \S\ref{sec:galprops_den_evo}, but find no significant correlations between them. Here, we discuss possible interpretations to explain our findings and how these will affect the current consensus on galaxy evolution, absent trends with $n_{\rm{e}}$, and limitations of our work.

\subsection{Evolving Density and Ionization Structure of Galaxies at $z>4$}
\label{sec:structure}

To explain our findings, we explore how density and ionization structures changes with $z$ \citep[e.g.][]{cataldi_etal_2025, scholte_etal_2025}. Specifically, we use the fact that the density from [\ion{O}{2}] and [\ion{S}{2}] only probes the densest clumps of low-ionization gas \citep{osterbrock_flather_1959, kennicutt_1984} to propose that 1) the spatial distribution of nebular gas and 2) the degree of ionization as probed by O$_{32}$ evolves with $z$, at fixed $M_\star$. In Figure \ref{fig:neb_model}, we visualize two galaxies with different low-ionization gas distributions. Each galaxy has the same volume and number of gas particles, but with different gas clumping properties.

\begin{figure}
\begin{center}
    \includegraphics[width=0.4\textwidth, trim=30 0 30 0,  clip=yes]{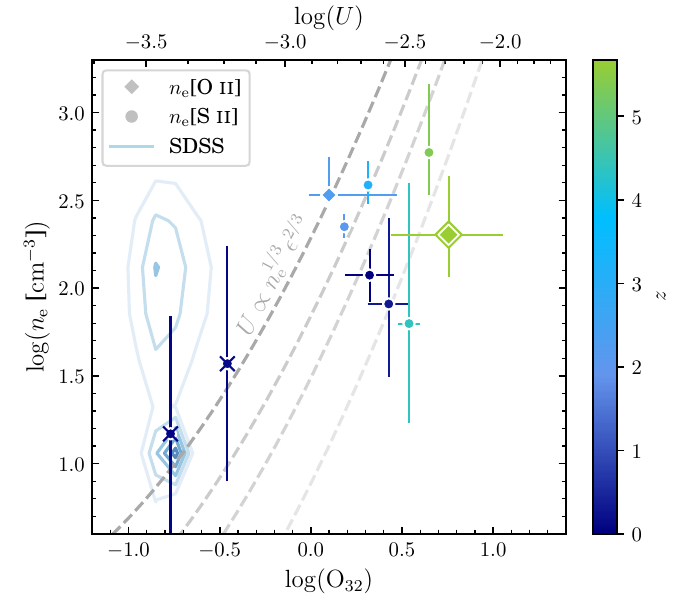}
        \caption{Similar to Figure \ref{fig:ne_O32_logU}, but without our empirical best-fit trend, and with the theoretical relation taking into account the variation of the volume filling factor $\epsilon$ over $0<\epsilon<1$, shown as increasingly lighter dashed grey lines. We infer that the observed deviation from the theoretical expectations of a radiation-bounded system is likely due to the fact that $\epsilon$ also evolves with redshift.}
\label{fig:ne_O32_logU_ff}
\end{center}
\end{figure}

We use the volume filling fraction $\epsilon$ to represent the degree of gas clumpiness in our model. We adopt the definition of $\epsilon$ as the density of emitting gas within the total \ion{H}{2} region volume \citep{kennicutt_1984}, which we approximate as:
\begin{equation}
    \epsilon\approx\frac{\langle n_{\rm{e}}\rangle^2}{n_{\rm{e, clump}}^2}.
\label{eq:13}
\end{equation}
Here, $\langle n_{\rm{e}}\rangle$ is the average electron density of gas, and $n_{\rm{e, clump}}$ is the electron density of an individual clump. As shown in our model (Fig. \ref{fig:neb_model}), $\epsilon$ ranges from $0\leq\epsilon\leq1$, with unity translating to a fully homogeneous medium, while $\epsilon<1$ translates to an increasingly clumpy medium. Recall that as long as $n_{\rm{ion}}/n_{\rm{H}}$ remains constant, emission-line strengths depend on $n_{\rm{e}}^2$ below $n_{\rm crit}$ of the given emission line, $n_{\rm{e, clump}}\sim n_{\rm{e}}$[\ion{O}{2}] and $n_{\rm{e}}$[\ion{S}{2}] for all values of $\epsilon$ \citep{osterbrock_flather_1959, osterbrock_1989, sanders_etal_2016, peimbert_peimbert_delgado-inglada_2017}. As $\epsilon\rightarrow0$, $n_{\rm{e, clump}}$ dominates the overall emission from the galaxy, biasing estimates of $n_{\rm{e}}$[\ion{O}{2}] and $n_{\rm{e}}$[\ion{S}{2}] to the densest clumps in the ISM. Therefore, the observed O$_{32}$ (and subsequently $U$) would also primarily be sensitive towards clumpy regions. We see evidence of this observationally in Figure \ref{fig:ne_O32_logU_ff}, where we overplot models with varying $\epsilon$ ($0<\epsilon<1$) (Eq. \ref{eq:3}) with our data. We are unable to directly constrain $\epsilon$ with our datasets since $\langle n_{\rm{e}}\rangle$ depends on the entire volume of unresolved \ion{H}{2} regions. However, increasing $\epsilon$ at fixed $n_{\rm{e}}$ manifests as an observed increase of $U$, as seen in Figure \ref{fig:ne_O32_logU_ff}.

To test our model, we compute $\epsilon$ using empirically-motivated prescriptions for $\langle n_{\rm{e}}\rangle$ and we adopt our best-fit double power-law evolution to serve as the evolution of $n_{\rm{e, clump}}$ (Eq. \ref{eq:11}). We derive lower limits for $\langle n_{\rm{e}}\rangle$ assuming that the entire star-forming volume of a galaxy is uniformly filled with ionized gas \citep{davies_etal_2021, reddy_etal_2023a}:
\begin{equation}
    \langle n_{\rm{e}}\rangle=\left[ \frac{L_{\rm{H\alpha}}}{\gamma_{\rm{H\alpha}} \cdot V} \right]^{1/2}=\left[ \frac{L_{\rm{H\alpha}}}{2\gamma_{\rm{H\alpha}} \cdot 2\pi r_e^2h} \right]^{1/2}.
\label{eq:14}
\end{equation}
Here, $\gamma_{\rm{H\alpha}}$ is the volume emissivity of H$\alpha$, and $h$ is the scale height, assuming the volume of the galaxy is a disk \citep{davies_etal_2021}. The first factor of 2 in the denominator accounts for $r_e$ being the radius encompassing half of the total $L_{\rm{H\alpha}}$ \citep[e.g.][]{reddy_etal_2023a}. To estimate $L_{\rm{H\alpha}}$, we use the star-forming main sequence of \citet{popesso_etal_2023} and the \citet{kennicutt_evans_2012} SFR-$L_{\rm{H\alpha}}$ relation (Eq. \ref{eq:7}), and for $r_e$, we use the size-redshift evolution of \citet{constantin_etal_2023}. We fix the stellar mass at $\log(M_\star)=8.9~M_\odot$, the median of all our samples with O$_{32}$. This estimates $L_{\rm{H\alpha}}$ and $V$ to derive lower limits for $\langle n_{\rm{e}}\rangle$ for galaxies across $z$.

\begin{figure*}
\begin{center}
    \includegraphics[width=\textwidth, trim=30 0 30 0,  clip=yes]{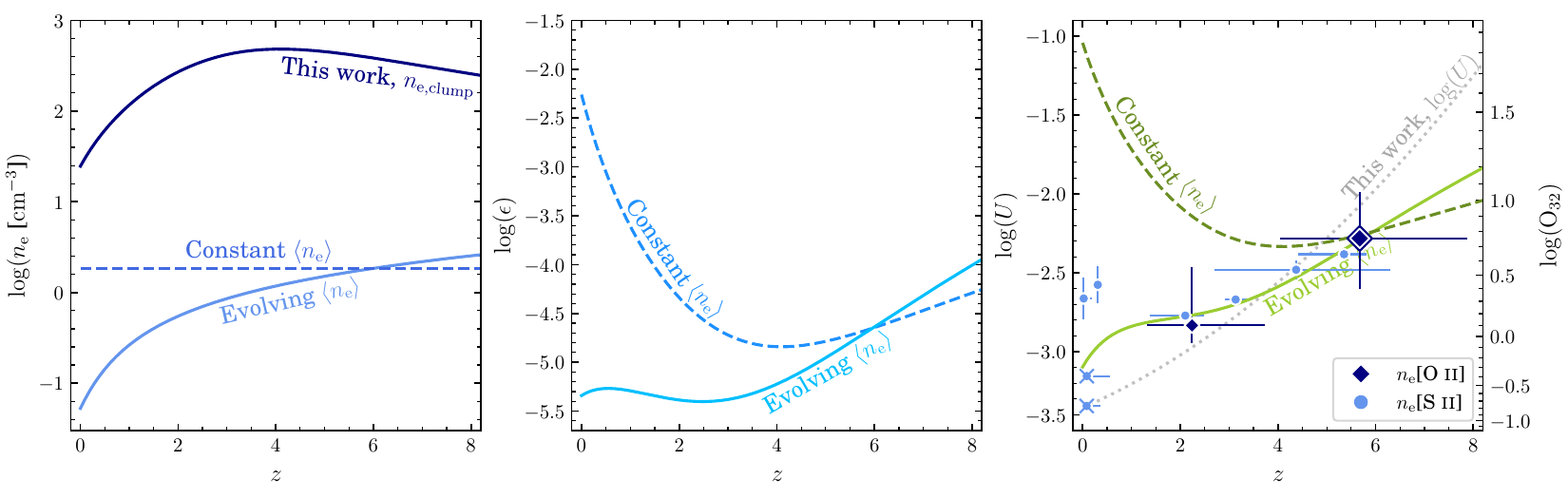}
        \caption{Empirically-motivated evolutionary trends for $n_{\rm{e, clump}}$ and $\langle n_{\rm{e}}\rangle$ (\textit{left}), $\epsilon$ (\textit{middle}), and $U$ (\textit{right}) based on our model shown in Figure \ref{fig:neb_model}. Here, we make predictions of how each of these properties evolve with $z$, assuming our best-fit law for $n_{\rm{e, clump}}$ is correct, and find that we are able to reproduce the observed behavior of $U$ for our collection of samples. For the predicted $\langle n_{\rm{e}}\rangle$ trend and subsequent dependent trends, $\langle n_{\rm{e}}\rangle$ over $z$ is evaluated at fixed $M_\star$ ($\log(M_\star)=8.9~M_\odot$). In the rightmost panel, the SDSS medians are demarcated with an ``x''. We find that our empirical predictions for the evolution of $U$ with $z$ match our combined datasets, supporting our theory that $\langle n_{\rm{e}}\rangle$ and $\epsilon$ must also evolve with $z$.}
\label{fig:ne_evo_theory}
\end{center}
\end{figure*}

In Figure \ref{fig:ne_evo_theory}, we show the three primary components of our empirically-based toy model. In the left panel, we display our best-fit double-power law we previously derived for our combined datasets across $z$ (\S\ref{sec:z_den_evo}, Eq. \ref{eq:11}) approximated as the evolution of $n_{\rm{e, clump}}$ (see Fig. \ref{fig:neb_model}). We also include the lower limit evolution of $\langle n_{\rm{e}}\rangle$ as discussed above (Eq. \ref{eq:14}). 

The middle panel shows the evolution of $\epsilon$ derived using Eq. \ref{eq:13}. We derive this trend for $\epsilon$ using the $n_{\rm{e, clump}}$ and $\langle n_{\rm{e}}\rangle$ trends shown in the left panel. The $\epsilon$ derived from the evolving $\langle n_{\rm e}\rangle$ exhibits a complex, non-monotonic evolution with redshift. At low-redshift ($z\sim 0$), \ion{H}{2} regions would be embedded in structurally complex, multiphase ISM, leading to significant density contrasts, low $\epsilon$ values, and a high-degree of clumpiness. Toward intermediate redshifts ($z\sim1-3$) increased gas fractions, turbulence and structurally complex star formation \citep{genzel_etl_2006, daddi_etal_2010, tacconi_etal_2010, tacconi_etal_2020} would establish another local minima in $\epsilon$. At higher redshifts ($z>4$), \ion{H}{2} regions would be more compact and less hierarchically structured, with the ionized gas dominated by a smaller number of dense regions rather than the large number observed at low-redshift. This reduction in internal structure would lead to an increase in $\epsilon$ at higher $z$. 

Finally, the right panel shows the evolution of $U$ derived using Eq. \ref{eq:3} with the evolution of $n_{\rm{e, clump}}$ and $\epsilon$. We also consider the fact that $Q$ likely evolves with $z$ (as it also depends on $L_{\rm{H\alpha}}$) and find that it does not affect the shape of the evolution of $U$. We also overplot $U$ for our data, which is derived using O$_{32}$ and the schema of \citet{berg_etal_2019}, in the right panel.

The data appear to visually match our predicted evolution of $U$, suggesting that both an evolving $n_{\rm{e, clump}}$ and $\langle n_{\rm{e}}\rangle$ may explain how $U$ evolves with $z$. For comparison, we also include a constant evolution of $\langle n_{\rm{e}}\rangle$, derived using a typical value of $\epsilon$ in the local universe \citep[$\epsilon\sim0.005$;][]{kennicutt_1984}, and our adopted $n_{\rm{e, clump}}$ at $z=0$ (Eqs. \ref{eq:10}, \ref{eq:11}). The assumption of a constant $\langle n_{\rm{e}}\rangle$ evolution fails to reproduce the observed evolution of $U$. Figure \ref{fig:ne_evo_theory} also shows that in order to reproduce the evolution of $U$ and $n_{\rm{e, clump}}$, assuming these trends hold true, $\epsilon$ must evolve with redshift.

However, we caution against drawing more definitive conclusions, particularly with regards to $\epsilon$, for a number of assumptions inherent within our model:
\begin{enumerate}
    \item[--] We assume a double power-law evolution for $n_{\rm{e, clump}}$, however the statistical analysis is unable to distinguish between this evolution and a flat (Eq. \ref{eq:10}) evolution. More observations at high redshift are needed to better establish whether or not either of these alternative evolutions is preferred.
    
    \item[--] Fixed stellar mass across cosmic time is not a realistic assumption \citep[e.g.][]{popesso_etal_2023, curti_etal_2023}
    
    \item[--] We assume $L_{\rm{H\alpha}}$ is emitted from the entire star-forming volume, however the line-emitting gas may not comprise the entire ionized volume of galaxies \citep[e.g.][]{kennicutt_1984, hunt_hirashita_2009, davies_etal_2021, reddy_etal_2023a}
    
    \item[--] The $n_{\rm{e, clump}}$ measurements only trace low-ionization gas across the limited density range of [\ion{O}{2}] and [\ion{S}{2}]
    
    \item[--] Our predictions of $\langle n_{\rm{e}}\rangle$ represent lower limits, and likely underestimate the true values of $\langle n_{\rm{e}}\rangle$, and the evolution with $z$, given that we rely on the H$\alpha$ size and luminosity \citep[e.g.][]{davies_etal_2021}
\end{enumerate}
Nevertheless, our inhomogeneous galaxy ISM model suggests that the nebular geometry of galaxies has significant evolution with redshift.

There are also theoretical predictions for why galaxies are clumpier and more inhomogeneous at high-$z$. Recently, work has been done using the MEGATRON \citep{katz_etal_2024, choustikov_etal_2025} simulations. They find that the ISM of high-$z$ MEGATRON galaxies on average are denser, more inhomogeneous, more metal-poor, subject to harder ionizing radiation fields, and have higher ionization parameters than low-$z$ galaxies \citep{choustikov_etal_2025}. These simulations agree with both our results and other \textit{JWST} studies \citep[e.g.][]{arellano-cordova_etal_2022, bunker_etal_2023, abdurrouf_etal_2024, sanders_etal_2024}.

Interestingly, MEGATRON finds that the O$_{32}$ ratio transitions from an ionization parameter estimate to a density estimate. [\ion{O}{2}]$\lambda\lambda3727,3730$ ÅÅ collisionally de-excite at a lower density than [\ion{O}{3}]$\lambda5008$ Å, which roughly corresponds to $\log$(O$_{32})>1.25$ in the simulations \citep{choustikov_etal_2025}. At this threshold, it is also believed that essentially all H gas is ionized, in which case O$_{32}$ will overestimate $U$ \citep[e.g.][]{brinchmann_etal_2008, reddy_etal_2023a}. We do not find median values above this threshold in this work, but note that if this is indeed the case overall, then we would expect to see the correlation between O$_{32}$ and $n_{\rm{e}}$ to continue. However, we note that MEGATRON simulations are primarily of high-$z$ galaxies, and caution against interpolating these findings to low-$z$ and local galaxy ISM density studies.

\subsection{Considerations of a Density-Bounded System}
\label{sec:den_bound}

We have discussed the likelihood of a radiation-bounded scenario (Eq. \ref{eq:3}) in explaining the evolution of $U$ and $n_{\rm{e}}$ in galaxies across redshift, but there remains the possibility that our results can be explained by a density-bounded system. In a density-bounded nebula, three conditions will be met; 1) the escape fraction of ionizing photons ($f_{\rm{esc}}$) will be nonzero, 2) the O$_{32}$ ratio will overestimate $U$, and 3) the theoretical relation (Eq. \ref{eq:3}) will change to $U\propto n_{\rm{e}}^{-1}$ \citep[e.g.][]{brinchmann_etal_2008, reddy_etal_2023b, reddy_etal_2023a}. Galaxies are assuredly not purely density-bounded, so as $f_{\rm{esc}}$ approaches unity there will be a transition from the radiation-bounded relation to the density-bounded relation. This suggests that the redshift evolution of $U$ and $n_{\rm e}$ could plausibly be driven by an evolution of $\langle f_{\rm esc}\rangle$.

It is also possible that some galaxies are better explained by a radiation-bounded system and others explained by a density-bounded system, and that some may be explained by some combination of both. All of these possibilities would cause the observed trend between $U$ and $n_{\rm{e}}$ to deviate from theoretical expectations. In addition, a nonzero $f_{\rm{esc}}$ could also be explained by increased clumpiness and/or inhomogeneities within the gas distribution of a system (i.e. $\epsilon<1$), which would translate to an inverse correlation between $\epsilon$ and $f_{\rm{esc}}$. It would be prudent to investigate whether or not there is such a connection, and it may be possible with a sample of galaxies with accurate measurements of multi-phase $n_{\rm{e}}$, $U$, $\epsilon$, and $f_{\rm{esc}}$. We plan to explore this further in a future paper, however with this current collection of galaxy samples, we cannot accurately constrain nor break the degeneracy between $\epsilon$ and $f_{\rm{esc}}$.

\subsection{Limitations}
\label{sec:limits}

Here we discuss a myriad of limitations and caveats to consider when interpreting the results of our study. We include discussions regarding biases arising from observations, as well as those from systematics that are inherent to ionized nebular gas studies.

\subsubsection{Observational Biases}
\label{sec:bias}

There is a concern that the NIRSpec/MSA sample does not represent a typical galaxy population at $z\sim5.6$: most of GO 1871 sources are fainter than a typical galaxy at the same redshift \citep[see Fig. 3 in][]{saldana-lopez_etal_2025}. The sample variance of the NIRSpec/MSA dataset is something we must consider when completing our analysis, but we find that when we account for this error in the fitting process, the fit parameters, $\chi^2_{\nu}$ values, and BICs all remain consistent within their respective uncertainties. We also test the impact of stacking on the median $n_{\rm{e}}$ by stacking the NIRSpec/MSA sample instead of taking the median $n_{\rm{e}}$ using our procedure outlined in \S\ref{sec:congress_fresco_samp}. We do this for both the density NIRSpec/MSA sample, and the full GO 1871 and GLASS sample with coverage of [\ion{O}{2}]. When we remeasure the stacked median $n_{\rm{e}}$[\ion{O}{2}] following the schema outlined in \S\ref{sec:dens}, and refit our trends to the median, we find that the fit parameters, $\chi^2_{\nu}$, and BIC values also remained consistent within their respective uncertainties. We also tested whether or not stacking the MSA spectra would allow us to measure a median $n_{\rm{e}}$[\ion{S}{2}], but it was within the uncertainty of the $n_{\rm{e}}$[\ion{S}{2}] of 1871-545 (see Fig. \ref{fig:go_S2}), the only galaxy in the NIRSpec/MSA sample with a [\ion{S}{2}] detection. We checked to see the profile after stacking the remaining galaxies with coverage of [\ion{S}{2}], and found that neither stacked doublet component passed the S/N $\geq3$ threshold. While we cannot quantify the completeness for the NIRSpec/MSA sample, it is important to note that this dataset was selected for the purpose of highlighting the need for more detailed, high spectral resolution studies of galaxies at early times. 

As for the NIRCam/grism datasets, the question of completeness is less of an issue, except for galaxies below the targeted luminosity limit. There is a possibility that these observations are missing a large fraction of galaxies that exist at these redshifts, but more in-depth observations that probe lower intrinsic luminosities are needed to determine how large of an effect this has on our results. In addition, we are unable to account for all galaxies within the NIRCam/grism sample that have densities beyond the low-density, critical density, and high-density boundaries, as we are unable to detect the [\ion{S}{2}] doublet in each individual galaxy included in the stack. Although we attempt to account for these uncertainties, this could affect our median density estimate, which would then alter the overall density evolution beyond the $z>4$ boundary. There is also a concern that the stacked [\ion{S}{2}] will be biased towards contributions from low-ionization sources, given these would have higher [\ion{S}{2}]/H$\alpha$ ratios and that the spectra are normalized by H$\alpha$. To check this, we search for correlations between [\ion{S}{2}]/H$\alpha$, O$_{32}$ and $L_{\rm{H \alpha}}$ in SDSS DR7 and DR16, and find that there is a shallow anticorrelation between [\ion{S}{2}]/H$\alpha$ and O$_{32}$, which would bias towards a lower $n_{\rm{e}}$[\ion{S}{2}] derived from an H$\alpha$-normalized stack. To test whether or not this is a significant effect, we perform the stacking procedure again for the NIRCam/grism galaxies but instead create two stacks of bright ($\log(L_{\rm H\alpha}/{\rm erg~s^{-1}})>42$) and faint ($\log(L_{\rm H\alpha}/{\rm erg~s^{-1}})>42$) galaxies. If there are differences in the relative line strengths and line profiles of [\ion{S}{2}], then this would change the median $n_{\rm{e}}$[\ion{S}{2}]. However, we see no changes to the [\ion{S}{2}] profiles that would affect $n_{\rm{e}}$[\ion{S}{2}], and conclude that this bias would have a negligible effect on the median $n_{\rm{e}}$ we estimate for the NIRCam/grism galaxies. This effect could also be present for the ancillary datasets we include that are stacks normalized by H$\alpha$, however we cannot directly quantify whether or not this would bias the median $n_{\rm{e}}$ we derive for those datasets.

\ion{H}{2} regions, and by extension the ISM of galaxies, are complex structures, and have been found to exhibit small-scale fluctuations and variations in both their temperature and density structures \citep{peimbert_1967, peimbert_2019, mendez-delgado_etal_2023a, mendez-delgado_etal_2023b}. The presence of spatially-unresolved density inhomogeneities can be problematic when making inferences on the overall physical conditions of ionized nebulae \citep{mendez-delgado_etal_2023b}. They find that the applicability of density diagnostics on a particular region of gas that houses density inhomogeneities is determined primarily by the sensitivity range inherent to the diagnostic. This is contrary to what is typically done for many ionized nebular gas studies, where the applicability of a given density diagnostic is typically determined by the ionization zone that corresponds to that gas region. This finding has widespread implications for multiphase studies of nebular regions, but for our work the [\ion{O}{2}] and [\ion{S}{2}] diagnostics have similar ionization energies and sensitivities, and thus are believed to trace similar regions of gas \citep[see \S\ref{sec:den_results},][]{mendez-delgado_etal_2023b}. However, undiagnosed density inhomogeneities may still bias our results. $R$[\ion{O}{2}] and $R$[\ion{S}{2}] underestimate electron densities by $\sim300$ cm$^{-3}$, even if densities are beyond the low-density limits \citep{mendez-delgado_etal_2023b}. If this is the case for our study, then this could alter our proposed evolutionary trend determined in \S\ref{sec:z_den_evo}. 

There are also numerous studies of galaxies dedicated to using rest-frame UV and FIR density-sensitive lines to compute electron densities, bringing multiphase nebular studies into focus at high-$z$ \citep[e.g.][]{mingozzi_etal_2022, topping_etal_2025, harikane_etal_2025b, martinez_etal_2025, peng_etal_2025a, usui_etal_2025}. Several have suggested an increasing evolution of $n_{\rm{e}}$ as measured with the high-ionization rest-frame UV tracer \ion{C}{3}] \citep{topping_etal_2025, harikane_etal_2025b, hayes_etal_2025, martinez_etal_2025}. \citet{choustikov_etal_2025} also finds that ISM densities inferred from rest-frame UV tracers are larger than those inferred from rest-frame optical tracers in MEGATRON galaxies, in agreement with observations \citep[e.g.][]{harikane_etal_2025b, hayes_etal_2025, usui_etal_2025}. If we extend this result to our work, assuming our low-ionization density evolution holds true, this would imply that as $n_{\rm{e}}$[\ion{O}{2}] and $n_{\rm{e}}$[\ion{S}{2}] decreases, $n_{\rm{e}}$\ion{C}{3}] increases at $z>4$. In other words, a larger fraction of gas is likely in high-density regions than low-density regions in high-$z$ galaxies (i.e., $\langle n_{\rm e} \rangle$ increases towards higher $z$), which is one of the key components of our toy model (Figure \ref{fig:neb_model}). However, to truly understand how ISM density and ionization structures evolve over time, we need more multiphase, panchromatic studies of galaxy ISM conditions, with multiple diagnostics detected in many galaxies across $z$.

\subsubsection{Systematic Uncertainties}
\label{sec:systematics}

In addition, underlying systematic uncertainties inherent to density diagnostics can significantly impact how we derive and interpret density measurements. There now have been several studies that account for systematics when looking at how $n_{\rm{e}}$ evolves in galaxies. Recently, \citet{peng_etal_2025a} test and quantify underlying systematics of several far-infrared (FIR) and optical density diagnostics, including [\ion{O}{2}] and [\ion{S}{2}]. They find that both [\ion{O}{2}] and [\ion{S}{2}] are found to estimate galaxy electron densities near the low-density limits for $\sim1/3-1/2$ of objects in their study. \citet{peng_etal_2025a} find that these tracers should not be used as galaxy-integrated densities without proper statistical treatment, and attribute low-density limit values to underestimated uncertainties in flux calibration, extraction, or measurements. Since density diagnostics are line ratios, these randomized errors can cause the measured value to fluctuate from what is expected from theory. We find that a similar fraction of NIRSpec/MSA galaxies have densities measured near or at the low-density limit (here, $R\sim1.5$ for both [\ion{O}{2}] and [\ion{S}{2}]), and we cannot rule out the possibility this is also true for the NIRCam/grism galaxies. 

Intrinsic errors derived from the ionic structure and excitation physics of the tracers themselves can further amplify these effects. At low- and high-density limits for a given tracer, ordinary error propagation techniques fail to account for the nonlinear relation between $R$ and $n_{\rm{e}}$ \citep{peng_etal_2025a}. We confirm and incorporate the presence of these effects when attempting to compute individual object densities, as well as sample median densities and uncertainties (see \S\ref{sec:dens}). This highlights the importance of dedicated investigations of systematics of all tracers of nebular conditions that have nonlinear relations with diagnostic emission line ratios. In summary, great care must be taken when deriving $n_{\rm{e}}$, regardless of whether it is for an individual \ion{H}{2} region, galaxy-integrated estimate, or sample/population average or median.

\section{Conclusions}
\label{sec:fin}

In this paper, we investigate the evolution of electron densities of galaxies at $4\leq z\leq 8$ using \textit{JWST} NIRSpec/MSA spectra from the GO 1871 and GLASS surveys, and NIRCam/grism spectra from the FRESCO and CONGRESS surveys, as well as a collection of galaxy samples across $0<z<8$ reported in the literature. This work demonstrates the power of large, complete, and highly detailed galaxy samples, and a homogeneous analysis schema, in studying galaxy densities and their evolution across cosmic epochs. We summarize our main findings in this section:

\begin{itemize}

  \item[--] We use \textit{JWST} NIRSpec/MSA and NIRCam/grism spectroscopy to target two samples of galaxies at $4\leq z\leq 8$ (Figure \ref{fig:samp_props}). We successfully detect the rest-frame optical density-sensitive emission line doublets [\ion{O}{2}] $\lambda\lambda$3727, 3730\AA\AA~(Figure \ref{fig:go_O2}) in the NIRSpec/MSA galaxies and [\ion{S}{2}] $\lambda\lambda$6718, 6733 \AA\AA~(Figure \ref{fig:congress_fresco_S2}) in the NIRCam/grism galaxy stacks.
  
  \item[--] We derive median $n_{\rm{e}}$[\ion{O}{2}] and $n_{\rm{e}}$[\ion{S}{2}] for the NIRSpec/MSA and NIRCam/grism samples, respectively (Figures \ref{fig:ne_R} and \ref{fig:ne_z_smol}). We find median electron densities of $n_{\rm{e}}$[\ion{S}{2}$]=463^{+281}_{-203}~{\rm{cm}}^{-3}$ for CONGRESS at $\langle z \rangle = 4.29$, $n_{\rm{e}}$[\ion{S}{2}$]=301^{+388}_{-206}~{\rm{cm}}^{-3}$ for FRESCO at $\langle z \rangle = 5.29$, and $n_{\rm{e}}$[\ion{O}{2}$]=202^{+145}_{-102}~{\rm{cm}}^{-3}$ for the combined NIRSpec/MSA sample at $\langle z \rangle = 5.68$.
  
  \item[--] We compile a supplementary collection of galaxy samples from the literature with median electron densities measured from [\ion{O}{2}] and [\ion{S}{2}]. With these combined datasets, we find that the median densities are inconsistent with an increasing single monotonically increasing power-law redshift evolution at $z \geq 3$ proposed by other studies in the literature (Figure \ref{fig:ne_z}), with extrapolated median densities relations from lower redshifts $\sim0.5-1$ dex higher than our results.
  
  \item[--] We find a correlation between median $n_{\rm{e}}$ and median O$_{32}$ (i.e. $U$) across redshift (Figure \ref{fig:ne_O32_logU}, Equation \ref{eq:12}), and note that it appears to be shallower than predicted by theoretical expectations (Eq. \ref{eq:3}) of a radiation-bounded nebula, assuming all other dependencies remain fixed. This suggests the observed evolution with $U$ is not solely driven by that of $n_{\rm{e}}$.
  
  \item[--] We do not find statistically significant trends between $n_{\rm{e}}$ and other global galaxy properties (Figure \ref{fig:ne_props}). However, this is likely due to the small number of medians and the limitations of each dataset.
  
  \item[--] We discuss a combination of an evolving density and ionization structure of the ISM of galaxies to explain how $n_{\rm{e}}$ and $U$ evolve with $z$. To visualize and quantify this, we propose an inhomogeneous ISM model (Figure \ref{fig:neb_model}) that uses the volume filling fraction $\epsilon$ to probe ionized gas clumpiness, which would have observable effects on the ionization state of the gas (e.g. Figure \ref{fig:ne_O32_logU_ff}). The toy model finds that the the average density smoothly rises with increasing redshift, while galaxies above $z \sim 3$ become less clumpy at $z > 3$ as morphologically complex galaxies become simpler, less developed star-forming regions. Our proposed model is able to explain the trends between $n_{\rm{e}}$ with $z$ and $U$, as well as the observed increasing evolution of $U$ with $z$ (Figure \ref{fig:ne_evo_theory}). It also supports the existence of an evolution of $\langle n_{\rm{e}}\rangle$ and $\epsilon$ with $z$. However, we caution against drawing more nuanced conclusions due to a number of inherent assumptions that went into developing our simplistic model.

  \item[--] We consider a number of limitations and caveats of our study, arising from both observational effects and physical processes. Notably, we discuss the effects that small-scale density inhomogeneities have on observed nebular properties, and the importance of multiphase, panchromatic nebular studies in probing a wider range of densities and ionization conditions. We then account for the effects of systematic uncertainties originating from density diagnostics and atomic physics. We also advocate for future works dedicated to quantifying systematic uncertainties in diagnostic tracers of nebular conditions.

\end{itemize}

% Our work highlights the potential of large ($N>100$), blind-search selected, high-quality ($R>1600$), and homogeneous galaxy samples at $z>4$ to study their ISM regions in exquisite detail. With \textit{JWST}, it is now possible to begin amassing similar samples that range across redshifts to piece together the evolution of these primitive structures into galaxies we see in the local universe. Our results advocate that high spectral and spatial resolution observations of the multiphase ISM across wavelength regimes are needed to adequately probe and understand the complexity of these structures, and understand how they influence galaxy evolution.

\begin{acknowledgments}

We are grateful for the support for programs 1871, 1895, and 3577 provided by NASA through a grant from the Space Telescope Science Institute, which is operated by the Association of Universities for Research in Astronomy, Inc., under NASA contract NAS 5-03127. This work is based in part on observations made with the NASA/ESA/CSA \textit{JWST}. The data were obtained from the Mikulski Archive for Space Telescopes at the Space Telescope Science Institute, which is operated by the Association of Universities for Research in Astronomy, Inc., under NASA contract NAS 5-03127 for \textit{JWST}. These observations are associated with programs 1324, 1871, 1895 and 3577. The specific observations analyzed can be accessed via \dataset[DOI: 10.17909/01dt-fm17]{https://doi.org/10.17909/01dt-fm17}.

ASL acknowledges support from Knut and Alice Wallenberg Foundation. TAH acknowledges support from an appointment to the NASA Postdoctoral Program at the NASA Goddard Space, administered by Oak Ridge Associated Universities under contract with NASA, as well as the University of Maryland Baltimore County and the Center for Space Sciences and Technology. NGG and YII and acknowledge support from the joint Ukrainian-Swiss project No. 224866. This work has received funding from the Swiss State Secretariat for Education, Research and Innovation (SERI) under contract number MB22.00072, as well as from the Swiss National Science Foundation (SNSF) through project grant 200020\_207349. The Cosmic Dawn Center (DAWN) is funded by the Danish National Research Foundation under grant DNRF140.

\end{acknowledgments}

\facilities{\textit{JWST} (NIRCam, NIRSpec)}

\software{\texttt{numpy} \citep[v.1.24.3;][]{numpy_etal_2020}, \texttt{scipy} \citep[v.1.13.1;][]{scipy_2020}, \texttt{astropy} \citet{astropy_2013, astropy_2018, astropy_2022}, \texttt{pyneb} \citep[v.1.1.24;][]{luridiana_morriset_shaw_2013, luridiana_morisset_shaw_2015, morisset_etal_2020}, \texttt{msaexp} \citep[v.0.8.4;][]{brammer_etal_2022}, \texttt{msafit} \citep{degraaff_etal_2024}, \texttt{grizli} \citep{brammer_etal_2018}, \texttt{EAZY} \citep{brammer_etal_2008}, \texttt{BAGPIPES} \citep{carnall_etal_2018}, \texttt{Prospector} \citep{johnson_etal_2019, johnson_etal_2021}, \texttt{pysersic} \citep{pasha_miller_2023}, \texttt{GALFIT} \citep{peng_etal_2002, peng_etal_2010}, \texttt{kendall} \citep{flury_etal_2022, flury_kendalltau}.}

\bibliography{jwst_den}{}
\bibliographystyle{aasjournal}

\end{document}